# Assessment of Neural Network Augmented Reynolds Averaged Navier Stokes Turbulence Model in Extrapolation Modes

Shanti Bhushan[1,*], Greg W. Burgreen[2], Wesley Brewer[3] and Ian D. Dettwiller [4]

[1] *Department of Mechanical Engineering, Mississippi State University; bhushan@me.msstate.edu*
[2] *Center for Advanced Vehicular Systems, Mississippi State University; greg.burgreen@msstate.edu*
[3] *Oak Ridge National Laboratory, Oak Ridge, TN, USA; brewerwh@ornl.gov*
[4] *Engineer Research and Development Center (ERDC); ian.d.dettwiller@usace.army.mil*

A machine-learned (ML) model is developed to enhance the accuracy of turbulence transport equations of Reynolds Averaged Navier Stokes (RANS) solver and applied for periodic hill test case, which involves complex flow regimes, such as attached boundary layer, shear-layer, and separation and reattachment. The accuracy of the model is investigated in extrapolation modes, i.e., the test case has much larger separation bubble and higher turbulence than the training cases. A parametric study is also performed to understand the effect of network hyperparameters on training and model accuracy and to quantify the uncertainty in model accuracy due to the non-deterministic nature of the neural network training. The study revealed that, for any network, less than optimal mini-batch size results in overfitting, and larger than optimal batch size reduces accuracy. Data clustering is found to be an efficient approach to prevent the machine-learned model from over-training on more prevalent flow regimes, and results in a model with similar accuracy using almost one-third of the training dataset. Feature importance analysis reveals that turbulence production is correlated with shear strain in the free-shear region, with shear strain and wall-distance and local velocity-based Reynolds number in the boundary layer regime, and with streamwise velocity gradient in the accelerating flow regime. The flow direction is found to be key in identifying flow separation and reattachment regime. Machine-learned models perform poorly in extrapolation mode, wherein the prediction shows less than 10% correlation with Direct Numerical Simulation (DNS). *A priori* tests reveal that model predictability improves significantly as the hill dataset is partially added during training in a partial extrapolation model, e.g., with the addition of only 5% of the hill data increases correlation with DNS to 80%. Such models also provide better turbulent kinetic energy (TKE) and shear stress predictions than RANS in *a posteriori* tests. Overall, ML model for TKE production is identified to be a reliable approach to enhance the predictive capability of RANS models. It is recommended that before a machine-learned model is applied for *a posteriori* tests, *a priori* tests should be performed to ensure that the model is not operating in full extrapolation mode and has reasonable accuracy.

---

[*] Author to whom correspondence should be addressed: *bhushan@me.msstate.edu*

# I. INTRODUCTION

The number of publications in the field of machine learning have seen an exponential rise in the last five years across all disciplines, and same is true for fluid mechanics (Fig. 1a) (based on bibliometric analysis from [1]). Machine learning is being applied for almost every aspect of fluid mechanics, and turbulence is at top of the list along with computational fluid dynamics (Fig. 1b). This is not surprising as turbulence modeling and simulation remains the largest source of uncertainty in computational fluid dynamics [2,3]. The following provides a review of the application of machine-learning (ML) for turbulent flow simulations.

The ML approach has been applied for several aspects of flow turbulence, such as: (1) to directly predict the turbulent flow field [4-10], such as the entire velocity field [4-6], scalar concentration in the wake [7], to estimate the initial flow field to speed-up the simulation [8], or to synthetically generate inflow turbulent flow field [9,10]; (2) to characterize the type of turbulent flow field, such as whether the flow is laminar, transitional or turbulent [11], type of turbulence – convective, wake or jet flow [12] or classify the turbulent structures [13], or analyze the local flow conditions to determine the most suitable turbulence model [14]; (3) for physics-informed neural networks (PINNs) to directly solve the Reynolds Averaged Navier-Stokes equations (RANS) for flow over a flat plate and airfoil [15]; (4) for assessment of the RANS turbulence modeling uncertainties [16-21]; and (5) for turbulence modeling. Turbulence modeling is of primary interest of this study, and in this area, prior research has been performed to: (a) evaluate RANS model coefficients; (b) obtain wall model for large eddy simulation (LES); (c) model subgrid stresses for LES; and (d) model Reynolds stresses for RANS. The research performed in each of the above categories are reviewed below focusing on input parameters, training approaches, training datasets and test cases, ML model validation and robustness tests, and feature importance analyses used in the studies.

For research area (a) involving the evaluation of RANS model coefficients, a vast amount of research has been done by Sandberg and co-authors [22-24], and Dwight and co-workers [25-28]. Sandberg and co-authors used Gene expression programming, a population-based optimization function, to estimate the coefficients of explicit algebraic Reynolds stresses and turbulent heat flux coefficients. The model coefficients were trained and applied to canonical test cases, such as duct flow and natural convention between heated plates, and to engineering applications, such as flow in coolant system and low-pressure turbines. Dwight and co-workers used symbolic regression to evaluate the coefficients of the non-linear component of the Reynolds stresses, and was applied to canonical cases, such as flow over periodic hill, wall mounted cube, and to the engineering application of prediction of multiple wind turbine constellation wake. Beetham and Capecelatro [29] also used symbolic regression to determine the coefficients of ten non-linear terms (invariants of rate-of-strain and rotation tensors) of an explicit algebraic RANS model. Sotgiu et al. [30] used deep neural networks (each having eight network layers with eight neurons per layer) to evaluate the coefficients of the non-linear algebraic stress terms for plane and corrugated channel flows. All these studies have reported that the ML augmented non-linear turbulence models performed better than the linear models and that the model coefficients worked well for unseen geometries, i.e., geometries which were not part of the training dataset.

For research area (b) involving the wall model of LES, Yang et al. [31] used plane channel flow direct numerical simulation (DNS) datasets, in particular velocity components, grid size and aspect ratio, and pressure gradient as input features to obtain a model for wall shear stress. Their model was trained using a neural network with four hidden layers, each containing five to ten neurons and applied to channel flow at a Reynolds number ($Re$) higher than those in the training set. Their study reported that inclusion of grid aspect ratio and pressure gradient did not improve results. Zangeneh [32] used DNS datasets of supersonic flow over a flat plate and an expansion-compression corner. Their input features included velocity, density, strain-rate, vorticity, pressure and temperature gradients, and grid size to obtain a model for the correction of wall stress and heat transfer for detached eddy simulation (hybrid RANS/LES) model. The model was trained using a Random Forest network with 300 trees and validated for cases with similar geometry as the training set but at higher $Re$.

For research area (c) involving LES subgrid stresses, stand-alone LES models have been developed for 2D turbulence [14,33-34], isotropic decaying turbulence for both incompressible and compressible flows [35-41], scalar flux in isotropic decaying turbulence [42] and for plane channel flow [43]. Most of these studies directly predicted the subgrid stress and heat fluxes, whereas Yuan et al. [36] used ML to predict unfiltered velocity which was then used to estimate the subgrid stresses. Vollant et al. [42] used ML to both directly predict subgrid stresses as well as model coefficients and reported that the latter performed better than the former. The isotropic turbulence studies used both first and second order derivatives of resolved velocity, temperature gradients for compressible flows, and grid scale to train the model. The data scaling used by Prakash et al. [41] is noteworthy as they transformed input and output variables into rate-of-strain eigen-frame to ensure that the variables are Galilean, rotation and reflection invariant. The channel flow studies compared different combinations of the input features involving velocity gradients, rate-of-strain and rotation tensors, and wall-distances. It was found that a combination of velocity gradient tensor and wall-distance was more reliable that using the rate-of-strain and rotation tensors. These studies mostly used deep neural networks with one or two hidden layers and 500 to 1000 neurons in each layer. Chen et al. [34] used a somewhat smaller network with three hidden layers and 20 neurons in each layer. Wang et al. [34] reported that models developed using neural networks performed better than those obtained using the random forest algorithm. These studies demonstrated the applicability of the model for unseen cases by applying them for different filter widths than the training case (i.e., coarser or finer grids than the training case). Zhou et al. [37] also introduced clipping of stresses to avoid backscatter of energy. Wang et al. [35] performed feature importance analysis and reported that subgrid stresses are strongly correlated with both first and second order velocity gradients and are weakly correlated with the velocities themselves.

For research area (d) involving Reynolds stresses for RANS, a few studies have directly inferred the Reynolds stresses [25,44-45], whereas the largest volume of work has been devoted to augmenting the existing RANS model. Kaandorp and Dwight [25] used a tensor basis random forest model to train for Reynolds stresses using DNS datasets for the periodic hill and converging-diverging nozzle. They used up to 17 input features that included five invariants of rate-of-strain and rotation tensors, wall-distance based Reynolds number, pressure-gradient parameters, turbulent kinetic energy (TKE), and ratio of TKE and dissipation. Their model was applied for curved and straight backward-facing step and square duct flows to demonstrate its applicability in extrapolation mode. To enhance the robustness of the model, the ML stresses and linear model stresses were combined using a blending parameter which was gradually increased from zero (purely linear model) to 0.8 (80% ML model and 20% linear model) as the simulation progressed. McConkey et al. [45] used deep neural network with 14 hidden layers and 30 neurons in each layer to train separate models for turbulent eddy viscosity and non-linear stresses, which were then combined to obtain the Reynolds stresses. Their study used Keras hyperparameter tuner [46] to obtain the neural network training parameters. Their model was trained using DNS datasets of a family of periodic hill geometries with different slopes using 32 input features including wall distance based Re, invariants of rate-of-strain tensor, rotation tensor, gradients of TKE, and gradients of pressure. The model was applied for slightly different periodic hill geometries than those used in training with slopes both within and outside of the training dataset. Their model was applied as fixed correction, i.e., the correction was estimated using input features estimated from a converged RANS solution, and non-negative constraints were imposed on Reynolds stresses to enhance robustness. Their study also performed SHapley Additive exPlanations (SHAP) analysis to evaluate the importance of input features on training. The SHAP analysis revealed that wall-distance based Re, ratio of mean strain timescale to a turbulent timescale, and stress anisotropy magnitude were the most important features. Fiore et al. [44] developed a model for coefficients of the turbulent heat flux components for low-Prandtl (Pr) (liquid metal) flows, while a physics based algebraic Reynolds stress model was used for turbulent stresses. Their model was trained using a neural network with six hidden layers using DNS datasets of channel, boundary layer, and backward-facing step flows. Their study used eight input features including TKE, dissipation, rate-of-strain tensor, rotation tensor, and temperature gradient. Their study reported some convergence issues in the far-field regions (where gradients are zero) during model training, which were resolved by introducing noise into the gradients. In addition, limiters were applied to avoid large values of the coefficients. Their ML model for heat flux was applied for unseen cases with both similar and completely different geometries (rod bundle case), where the latter represents model testing in extrapolation mode. Their study reported that the performance of the ML model depended strongly on the mean flow field, which was solved using a physics-based model. Thus, their model was not completely tested in the extrapolation mode.

The studies that focused on RANS model augmentation can be further subdivided into three categories: (i) direct estimation of the turbulent eddy viscosity ($v_T$) for linear models [47-55], (ii) correction terms for the linear models [56-63], and (iii) enhancement of the accuracy of the turbulence transport equations used in linear models [64-68]. Studies in category (i) have been applied for both incompressible [47-50] and compressible flows [51-54]. They have either used neural networks with two to six hidden layers with 20 to 40 neurons per layer or random forest models with three to four hidden layers with 64 to 128 trees. Some studies performed optimization of the network size or the training parameters. For example, Marioni et al. [47] used a random search to optimize hyperparameters, whereas studies [48] and [50] performed a parametric study. These studies mostly used five to twelve input features based on magnitude of rate-of-strain, vorticity magnitude, wall distance based Re, pressure gradients, velocity direction, and entropy for compressible flow. One exception is the study by Liu et al. [50] that used terms of the turbulence transport equation as input features. Several of the studies determined that it is critical that machine learned models are trained using datasets from regions with different flow physics to ensure robustness. For this purpose, Zhu et al. [51, 52] developed separate models for the inner layer, wake layer and the far-field. Marioni et al. [47] used clustering of data to put equal importance to different flow regimes. Studies [51-53] introduced averaging of $v_t$ over neighboring cells to enhance numerical stability. They also used feature importance to obtain an optimal set of features for training and reported that vorticity was most important followed by wall-distance based *Re*. Their study identified that pressure gradient was not important, in contradiction with the findings of Yang et al. [65]. The robustness of their model was demonstrated by applying it for higher Re or Mach numbers (*Ma*) or slightly different geometries. Overall, the studies have reported very encouraging results using ML models when compared with physics-based linear model counterparts. Some studies identified additional advantages of the ML approach, e.g., Maulik et al. [49] reported that ML $v_T$ estimation sped up simulation time by 100% compared to the two-equation RANS model. Song et al. [54] reported that ML turbulence model had more lenient near-wall $y^+$ requirements compared physics-based models, and thus required smaller grid sizes and lower computational costs.

Studies in category (ii) involving correction of terms for linear models have either proposed to use corrections to shear stresses ($\Delta\tau$) over the linear model prediction [56-58,60-61] or to use corrections to turbulent eddy viscosity ($\Delta v_T$) [62-63]. The models have been developed for both k-ε/k-ω and SA turbulence models and have been mostly developed for incompressible separated flows, except for Wang et al. [58] which extended the Wang et al. [56] model for compressible flows, and He et al. [63] performed simulation for shock induced separation. The models have been developed using both Random Forests with 100 to 300 trees and neural networks with two to four hidden layers with 30 to 128 neurons in each layer. The choice of input parameters varied considerably between the studies from eight features used by Volpiani et al. [62] to 50 features used by [58,59]. The input features typically included rate-of-strain tensor, rotation tensor, second invariant of rate-of-strain tensor $Q$, pressure gradients, turbulence modeling terms, such as ratio of production, diffusion, and destruction terms, wall distance based Re, flow direction, and higher order invariants of rate-of-strain and rotation tensors. He et al. [63] concluded that dropping redundant input features can reduce the complexity of the ML model with limited sacrifice in accuracy. They used SHAP analysis to understand the role of flow features on turbulence. It was reported that shear enhances turbulent fluctuation whereas rotation reduces it, consistent with Reynolds stress transport equations. Further, turbulent fluctuations are enhanced when the pressure stress dominates over the shear stress, and vice versa. Both Wang et al. [56] and Yin et al. [60] identified wall distance based *Re* to be their most important feature followed by turbulent eddy to molecular viscosity ratio. They also reported that their machine learning frameworks required strict requirements on grid smoothness. The robustness of the model in extrapolation model was mostly demonstrated for higher *Re/Ma* and slightly different geometry. He et al. [63] also recommended training separate models for near- and far-wall zones to enhance the robustness of the ML model.

The studies using field inversion approach fall into category (iii) that involve the enhancement of accuracy of the turbulence transport equations [64-68]. In this approach, a correction to the SA turbulence model production term is derived to enhance the accuracy of the linear model. These models are derived using six input features: vorticity magnitude, ratio of turbulent and molecular viscosity, ratio of rate-of-strain and vorticity magnitude, shear stress, ratio of turbulence production and dissipation, and $f_d$, which are the terms of the SA model $\nu_T$ transport equation. These models have been developed using deep neural networks with two to four hidden layers, and 16 to 128 neurons per layer, and applied to flows over an airflow at an angle of attack [65], an airfoil with ice accretion [68], and a turbine blade [67]. These models have been tested for unseen geometries with limited success, e.g., Yan et al. [68] reported that ML model predictions deteriorated for larger angles of attack. Yang and Xiao [66] extended the field inversion approach to estimate the multiplier for transition time scale to modify the first mode transition time scale and applied their model for transition over an airfoil surface. They also compared accuracy of models trained using a neural network with two hidden layers with 50 neurons each and those trained using a Random Forest model with 200 trees. The study reported that former performed somewhat better than the latter.

Overall, based on review of the literature, the following five key conclusions can be drawn regarding ML turbulence modeling, training, and validation approaches.

(1) **ML modeling**. Most studies have focused on RANS modeling, in particular, on enhancing existing RANS models by incorporating an ML model for stress non-linearity. Emphasis on RANS modeling is understandable, since it remains the model of choice for engineering application and imposes less stringent modeling requirements than LES that requires the entire turbulence regime to be modeled. In addition, using an ML model for stress anisotropy is also understandable since linear (isotropic) models are widely used based on their simplicity and robustness, yet turbulence is expected to be anisotropic in engineering applications. Limited research has been performed for direct RANS modeling since this involves a strong dependence on an ML model. Since ML models do not satisfy spatial and temporal continuity, they can introduce numerical instability [69]. Turbulence transport equation enhancement has not been applied that frequently, even though the field inversion approach was one of the pioneering approaches. In this study, we investigate the turbulence transport equation enhancement approach to directly address the limitations of RANS models for engineering applications for which the main flow quantities of interest are mean velocity and TKE and for which prediction errors for TKE are significantly large compared to the mean velocity [70]. Thus, ML models that directly improve TKE predictions provide a valuable contribution. Further, since the ML model does not affect the momentum term directly, the lack of spatial and temporal smoothness of ML model is mitigated as a potential source of numerical instability.

(2) **Input parameters**. A substantial amount of work has been done in identifying appropriate sets of input features for RANS model development and enhancement. Several studies suggest more features are better, whereas other have shown that larger sets of parameters introduce addition sources of error because of uncertainties in the model predictions due to numerical or grid convergence issues. Non-dimensionalization of flow features has been recommended, but scaling of the data has been most frequently used to transform them to a range of -1 to 1. The authors in a previous study [69] advocate for limited large-scale features that can be easily computed using simulations, and non-dimensionalizing them to obtain physical quantities, e.g., training using $Re_T = U\delta/\nu$ as a physical feature rather than training using the independent fields of kinematic viscosity ($\nu$), local velocity ($U$) and wall-distance ($\delta$). Further, it has been concluded that inclusion of higher order functions of an input variables does not necessarily improve predictions; however, they can be beneficial as weighting functions especially in the regimes where input datasets overlap

(3) **Training approach.** Studies have used both Random Forest and deep neural networks, and the latter is more prevalent than the former. Overall, studies have used small networks with four to six hidden layers with 10s to around 100 neurons in each layer. Some studies have performed optimization for the network and/or training parameters; however, no concrete conclusions have been drawn regarding the trade-off between training parameters and the model accuracy. More importantly, the uncertainties associated with the non-deterministic nature of the neural-network training has not been quantified. In this study, the role of the network and training parameters and associated uncertainties are investigated in detail.

(4) **ML model robustness**. Studies have been mostly applied for steady flows (or quasi steady flow for LES), and limited studies have considered separated flows. The robustness of the ML model in extrapolation mode have been limited to inferencing using higher *Re* and slightly different geometries. Studies have noted that universal models that works well for different flow regimes are difficult to train and have advocated the use of separately trained models for each flow regime. The research of [69] demonstrated that data clustering is an efficient way to ensure that a ML model incorporates different flow features and is not skewed towards a dominant flow feature. In this study, the predictive capability of ML model in extrapolation mode for separating-reattaching flow is tested, and the effect of data clustering on ML model accuracy is investigated in detail.

(5) **Flow physics analysis**. ML can provide feature importance estimates, such as using SHAP analyses, which can help identify most important input feature correlations with turbulence quantities and help quantify correlation strengths between turbulence and mean flow features. Only a few studies have used SHAP to perform such analyses. In this study, SHAP analyses are performed in the different regimes of the flow, e.g., shear layer, flow recirculation, reattachment, boundary layer and accelerating flow regimes, to critically evaluate correlation strengths between turbulence production and large-scale mean flow features.

This research builds upon the authors past study in [69], which investigated the predictive capabilities of ML RANS turbulence models (direct stress inference approach) for inner boundary predictions. To achieve this, a neural network was trained for shear stress using DNS datasets for steady and oscillating plane channel flows. The trained ML model was coupled with a pseudo-spectral solver and applied for the same test cases, but for different *Re* than those used during training. A physics-informed machine learned (PIML) model was also investigated for the plane channel case, for which the governing equations could be simplified to a one-dimensional equation and could be easily implemented during training. Overall, the machine learned turbulence model provided reasonable predictions, and the inclusion of the PIML approach enabled the ML model to converge to a correct solution from ill-posed conditions.

This study focuses on evaluating the abilities of machine-learned models in extrapolation mode. For this purpose, machine learned models are trained to predict the turbulent kinetic energy production using DNS/LES flows involving separation and reattachment and applied to a test case that has a significantly larger flow separation region and higher turbulence than those cases in the training set. The ML models are trained including different percentages of the test case data, ranging from 0% (full extrapolation mode) to 15% (partial extrapolation mode), and its effect on model accuracy is investigated for both *a priori* and *a posteriori* tests. The study also investigates: (1) the relationship of mini-batch and network sizes to training/validation residuals, (2) the uncertainty in ML model accuracy because of the non-deterministic nature of neural network training, (3) the effect of data scaling and data clustering on the ML model accuracy, and (4) the applicability of ML model in extrapolation mode.

Section II provides a summary of the validation and training dataset used in the study. Section III reports how grid refinement is related to the accuracy of the linear RANS model and describes the framework of the proposed ML model. Section IV provides a summary of the salient points of the deep neural networks of this study. Section V performs a sensitivity analysis of ML model on network and training parameters. Section VI performs a feature importance study. Section VII reports on validation of the ML model in *a priori* and *a posteriori* tests, and key conclusions are drawn in Section VIII.

## II. VALIDATION AND TRAINING DATASETS

In this study, a DNS/LES database of nine different incompressible separated flow cases have been curated from the NASA LARC website [71], as summarized in Table 1. The test cases include: (1) LES of flow over a bump with five different heights 0.02m (h20), 0.26m (h26), 0.031m (h31), 0.038m (h38) to 0.042m (h42), and flow over a curved backward-facing step (CBFS) and (2) three DNS test cases: flow through a converging-diverging nozzle (CDNoz), flow over a hump (Hump), and flow over a periodic hill (Hill). As evident from the streamlines and contours in Table 1, the h20 case does not involve any separation, whereas the Hill case shows the largest separation bubble, and the rest of the cases show separation bubbles with sizes that lie in between the h20 and Hill cases. The test cases involve several distinct flow physics regimes, e.g., boundary layer flow (mostly on the top wall), the separated flow region, flow reattachment, boundary layer flow with adverse pressure gradient (upstream of the obstacle), and free-shear region when the flow moves freely past the top of the obstacle.

To ensure that all the test cases are physically consistent, all spatial dimensions were normalized by the obstacle height, and the flow variables were normalized using obstacle height as the length scale and free-stream or channel-centerline-velocity as the velocity scale. Figure 2 shows a box and whisker plot of some of the key flow features (which will be used in this study either as input or output features). Features include rate-of-strain components $D_{uu} = \frac{\partial u}{\partial x}$ and $D_{uv} = \frac{1}{2}\left(\frac{\partial u}{\partial y} + \frac{\partial v}{\partial x}\right)$, rotation component $D_{uv} \frac{1}{2}\left(\frac{\partial u}{\partial y} - \frac{\partial v}{\partial x}\right)$, flow direction $\left(\frac{\bar{u}}{|u|}, \frac{\bar{v}}{|u|}\right)$, TKE and TKE productions ($P_K$), and wall-distance and local velocity based Reynolds number $Re_l = u\delta/\nu$. The box region shows the range of distribution of the 50% of the datasets, whereas the top and bottom ends of the whisker show the median of the 1st quartile and 4th quartile of the datasets, respectively. The Hill case involves lowest $Re_l \leq 1.53\times10^4$, followed by CDNoz with $Re_l \leq 2.3\times10^4$. The Hump case has largest $Re_l \leq 1.2\times10^6$. The remaining cases have $Re_l \sim O(10^5)$. In Fig. 2, Note that the plot is shown for $Re_l^{1/9}$ rather than for $Re_l$ because the power scaling of $Re$ provides a better distribution (wider spread) of the variable, which is more conducive for ML training. This aspect will be discussed in detail in Section V. One can observe that the mean flow variables for the Hill case lie within the range of the other eight cases, but the ranges of turbulence quantities are very different. Thus, an application of a ML turbulence model developed using Bump, CBFS CDNoz and Hump cases to predict the outlying Hill case will test its predictive capability in an extrapolation mode.

## III. ASSESSMENT OF RANS MODEL FOR PERIODIC HILL CASE

### A. Governing Equations and Turbulence Modeling

Before focusing on the development of the ML turbulence model for the Hill case, the limitations of a commonly used physics-based turbulence model are quantified for this case. For this purpose, RANS simulations are performed using OpenFOAM [72]. The governing equations for this case are the incompressible boundary layer equations, which involve enforcing both conservation of mass and momentum:

$$\frac{\partial \bar{u}_i}{\partial x_i} = 0 \quad \text{1(a)}$$

$$\frac{\partial \bar{u}_i}{\partial t} + \frac{\partial}{\partial x_j}(\bar{u}_i \bar{u}_j) = -\frac{1}{\rho}\frac{\partial \bar{p}}{\partial x_i} + \nu \frac{\partial^2 \bar{u}_i}{\partial x_j \partial x_j} - \frac{\partial \tau_{ij}}{\partial x_j} \quad \text{1(b)}$$

where, $i = 1, 2$ and 3 represent the $X$, $Y$ and $Z$ directions, respectively. $\bar{p}$ is the static pressure, and $\nu$ is the kinematic viscosity. The overbar $(\bar{\cdot})$ represents the ensemble averaged flow variables. The term $\tau_{ij}$ on the right-hand side of Eq. 1(b) are the turbulent stresses:

$$\tau_{ij} = -\overline{u_i' u_j'} \quad \text{1(d)}$$

These terms are modeled using the Boussinesq eddy viscosity approximation as:

$$\tau_{ij} = 2\nu_T S_{ij} \quad \text{1(e)}$$

where, $\nu_T$ is the turbulent eddy viscosity, and $S_{ij}$ is the rate-of-strain tensor. The $\nu_T$ is obtained using $k - \omega$ SST model [73], which requires the solution of two additional transport equations for turbulent kinetic energy ($k$) and specific dissipation ($\omega$), as below:

$$\underbrace{\frac{\partial k}{\partial t} + \frac{\partial (\bar{u}_j k)}{\partial x_j}}_{Advection} = \underbrace{P_K}_{Production} - \underbrace{\beta^* \omega k}_{Dissipation, \varepsilon} + \underbrace{\frac{\partial}{\partial x_j}\left[(\nu + \sigma_k \nu_t)\frac{\partial k}{\partial x_j}\right]}_{Diffusion} \quad \text{(2a)}$$

$$\frac{\partial \omega}{\partial t} + \frac{\partial (\bar{u}_j \omega)}{\partial x_j} = \underbrace{\frac{\gamma}{\nu_t} P_K}_{Production} - \underbrace{\beta \omega^2}_{Dissipation} + \underbrace{\frac{\partial}{\partial x_j}\left[(\nu + \sigma_\omega \nu_t)\frac{\partial \omega}{\partial x_j}\right]}_{Diffusion}$$

$$+ \underbrace{2(1 - F_1)\frac{\sigma_{\omega,2}}{\omega}\frac{\partial k}{\partial x_j}\frac{\partial \omega}{\partial x_j}}_{Cross-diffusion} \quad \text{2(b)}$$

The transport equation involves advection, production, dissipation and diffusion terms (both due to molecular and turbulent viscosities and cross-diffusion). A limiter is applied to the production term to avoid turbulence generation in flow stagnation regions as below:

$$P_K = min\left(\tau_{ij}\frac{\partial \bar{u}_j}{\partial x_j}, 20\beta^*\omega k\right) \qquad 2(c)$$

The turbulent viscosity is defined as:

$$\nu_t = \frac{a_1 k}{max(a_1\omega, \Omega F_2)} \qquad 2(d)$$

Readers are referred to [73] for details of the blending functions $F_1$ and $F_2$ and model coefficients.

### B. Simulation Setup and RANS Predictions

OpenFOAM provides a suite of numerical methods for CFD. In a previous study [72], the authors have identified that among the numerical methods available in *OpenFOAM*, the most appropriate models for RANS are the 2$^{nd}$-order implicit unbounded backward scheme for time discretization, and the 2$^{nd}$-order linear upwind bounded scheme for both convection, divergence, Laplacian and the surface-normal gradient terms. The 2$^{nd}$ order linear upwind bounded scheme is also used for the $k$ and $\omega$ equations. A combined *PISO-SIMPLE* algorithm called *PIMPLE* is used to satisfy mass conservation (via the pressure Poisson equation); this scheme allows use of a larger time step than *SIMPLE* and is less computationally expensive than *PISO*. Note that since the flow is expected to converge to a steady state, only two inner iterations of the predictor-corrector step are used.

The simulations are performed using a 3D domain with streamwise direction extent of X = [0, 9h], wall normal extent of Y = [0, 0.3h] and spanwise extent of Z = [0, 4.5h], where $h$ is the hill height selected to be unity, as shown in Fig. 3. A no-slip wall boundary condition is used for both the bottom and top walls, and a periodic boundary condition is used for the inlet and outlet pair and for the side domain pair. A body forcing is applied along the streamwise direction to match the friction and pressure drag from the top and bottom walls reported in the DNS. Even though the domain is 3D, a very coarse grid with two cells is used along the Z direction to maintain 2D flow. The simulations are performed on four systematically refined grids using a grid refinement ratio of $r_G = 2^{1/4}$. The four grids consist of grid points of sizes 106×70 (Coarsest), 125×85 (Coarse), 150×100 (Medium) and 180×120 (Fine) along X×Y directions. The grids are refined both on the top and bottom wall to obtain a near-wall grid resolution of $y^+ < 1$.

The RANS predictions obtained on all the four grids are compared with DNS in Figs. 4 and 5. Figure 4 compares the contour plots of streamwise velocity ($U$) along with flow streamlines and field variables of TKE, shear stress ($\tau_{uv} = \overline{u'v'}$) and $P_K$, which provides a qualitative assessment of the prediction. The streamlines show the separation bubble length and help in the assessment of the flow separation predictions. The values of TKE, $\tau_{uv}$ and $P_K$ help in the assessment of the turbulence predictions. Figure 5 compares the profiles of the mean flow and turbulence quantities at several cross-sections from X=1 to X=8, which is used for the grid convergence analysis and quantitative assessment of the RANS predictions.

The grid convergence analysis is performed for two different grid triplets ($S_1$, $S_2$, $S_3$) :: (Medium, Coarse, Coarsest), and ($S_1$, $S_2$, $S_3$) :: (Fine, Medium, Coarse). For this purpose, the solution convergence ratio R = ($S_1$-$S_2$)/($S_2$-$S_3$) was estimated for the velocity and TKE profiles at several cross-sections in the domain. Note that $0 < R < 1$ suggests a monotonic convergence of solution, $-1 < R < 0$ suggests an oscillatory convergence, and $R > 1$ and $R < -1$ suggest monotonic and oscillatory divergence, respectively. The grids are deemed to be reasonable when the solutions show monotonic convergence with small R, i.e., solutions are approaching an asymptotic range. The convergence study showed an average $R = 1.8$ and 0.73 for U and TKE, respectively, on the (Medium, Coarse, Coarsest) grid triplet, and an average $R = 0.6$ and 0.4 for $U$ and $TKE$, respectively, on the (Fine, Medium, Coarse) grid triplet. Thus, the Coarsest grid is too far from the asymptotic range, but both Medium and Fine grids seem to be reasonable for the study. Considering the computational cost and marginal difference in solutions between the Medium and Fine grids, the medium grid was selected for simulations in rest of the study.

Comparison of the RANS solution with DNS shows that the former significantly overpredicts the length of the flow separation region, i.e., $X_R = 7h$ in RANS compared to $5.2h$ in DNS. The velocity predictions are reasonable, and the averaged prediction error is estimated to be < 6%. The prediction towards the top wall is significantly better than those on the lower wall, and the largest errors are towards the inflow and in the reattachment regions. The turbulence quantities show significantly large errors, and the peak values are under predicted by 65%. The RANS simulations primarily fail to predict the high TKE in the shear layer above the separation bubble.

Overall, the results suggest that limitations of the k-ω SST model can be traced back to significantly lower shear stress, which is due to lower turbulent eddy viscosity or TKE predictions. The underprediction in TKE can be traced back to significantly lower TKE production. The proposed ML model focuses on addressing this limitation in TKE production.

### C. Proposed Neural Network Enhanced Turbulence Model

In this study a neural network is used to train a model for turbulence production ($P_k$). The model is trained using six input features: velocity direction $\frac{\bar{u}}{|u|}, \frac{\bar{v}}{|u|}$, rate-of-strain components $D_{uu}, D_{uv}$ and the rotation component $W_{uv}$ and $Re_l$. Note that rate-of-strain tensor component $D_{vv}$ is not used, as this component is same as $-D_{uu}$ (from mass conservation equation). The neural network training essentially provides a regression map of:

$$P_K = f\left(\frac{\bar{u}}{|u|}, \frac{\bar{v}}{|u|}, D_{uu}, D_{uv}, W_{uv}, Re_l\right) \tag{3}$$

During the CFD simulation, the machine-learned regression map is queried each time step and for each grid cell using the local input feature set, and the $P_K$ value provided by the model is used in the equation 2(a) and 2(b) instead of Eq. 2(c). Note that the maximum production limiter $20\beta^*\omega k$ is also applied for the ML model output. However, no limiter is used for the minimum. The physics-based model does not allow negative production, as it may result in negative TKE, which is unphysical. However, $P_K$ can be negative, as observed in the accelerating flow in the uphill region in Fig. 4 (DNS results). ML models are trained using DNS dataset which includes negative P_K, and this physically correct behavior is retained in this study. The coupling of the ML model with CFD solver is depicted in Fig. 6.

### IV. FUNDAMENTALS OF DEEP NEURAL NETWORK

The basic framework of a deep learning neural network is depicted in Figure 7, inspired from LeCun et al. [74]. A deep neural network consists of various layers, including an input layer comprising of input features (six herein), an output layer comprising of the required output features (one herein), and multiple hidden layers comprising of user specified features units (or neurons). The feature units in each hidden layer are obtained using linear combination of feature units from the previous layer and passing it through an activation function. For example, at first intermediate feature units $z_i^{(p)}$, $i = 1...M$ in the $p^{th}$ hidden layer is obtained using weighted linear combination of the feature units $y_j^{(p-1)}$, $j = 1...N$ in the $p-1^{th}$ layer, as below:

$$\begin{bmatrix} z_1^{(p)} \\ .. \\ ... \\ z_M^{(p)} \end{bmatrix} = \underbrace{\begin{bmatrix} w_{11}^{(p)} & \cdots & w_{1N}^{(p)} \\ \vdots & \ddots & \vdots \\ w_{M1}^{(p)} & \cdots & w_{MN}^{(p)} \end{bmatrix}}_{w_{ij}^{(p)}} \begin{bmatrix} y_1^{(p-1)} \\ .. \\ ... \\ y_N^{(p-1)} \end{bmatrix} \tag{4}$$

In the above equation, $w_{ij}^{(p)}$ are the unknown weights that need to be estimated. The weights are positive and normalized such that $\sum w_{ij}^{(p)} = 1$. Then the feature units in the $p^{th}$ layer are obtained from the intermediate feature units using a pre-defined non-linear analytic activation function, $f(..)$:

$$y_i^{(p)} = f(z_i^{(p)}) \tag{5}$$

The activation function introduces non-linearity in the network design and should be at least first-order differentiable. The most common activation functions are *ReLU* (rectified linear unit), Hyperbolic tangent, Sigmoid, and Logistic. *ReLU* function provides a linear dependency on the input features between 0 and 1, whereas the other functions modulate them between -1 and 1. Thus, based on the choice of the activation function, the training dataset should be scaled to be between [0,1] or [-1,1]. Note that an activation function is not used to obtain the output layer feature unit.

The unknown weight matrix in each layer is estimated using a backpropagation approach. For this, the network can be first initialized with constant zero weight for each layer, and errors in the network prediction are obtained by comparing value of the output feature ($y_O$) with the training dataset true values ($T_O$) using a user specified analytic cost function, such as $L_2$ norm:

$$E = \sum_{i=1,B}(y_{O,i} - T_{O,i})^2 \tag{6}$$

where the summation is over the data mini-batch size (*B*) used for training. The weights in each layer are adjusted as:

$$w_{ij}^{(p)}\big|_{it+1} = w_{ij}^{(p)}\big|_{it} - \alpha \frac{\partial E}{\partial w_{ij}^{(p)}} \tag{7}$$

where $\alpha$ is the user specified learning rate, and subscript '*it*' represents the iteration level. Note that "iterations" are the counters for the steps of passing a mini-batch of data (a subset of entire dataset) forward and backwards through the network, whereas "epoch" is a counter representing a single pass of the entire dataset forward and backward through the network. Thus, epochs are not same as iterations, unless the batch size enfolds the entire dataset in a single pass. The unknowns in the above equation are the first-order derivatives of the errors with respect to the weights, the second term on the right, which can be expressed as:

$$\frac{\partial E}{\partial w_{ij}^{(p)}} = \frac{\partial E}{\partial z_i^{(p)}} \frac{\partial z_i^{(p)}}{\partial w_{ij}^{(p)}} = y_j^{(p-1)} \frac{\partial E}{\partial z_i^{(p)}} \tag{8}$$

The derivatives of the errors with respect to the input features are:

$$\frac{\partial E}{\partial z_i^{(p)}} = \frac{\partial E}{\partial y_i^{(p)}} \frac{\partial y_i^{(p)}}{\partial z_i^{(p)}} = f'(z_i^{(p)}) \frac{\partial E}{\partial y_i^{(p)}} \tag{9}$$

The calculation of the last term on the RHS (derivative of error with respect to the layer feature) is computed based on information from the layer ahead, i.e., input features of a layer are linearly related to the output features of the previous layer as shown in Eq. (4), thus:

$$\frac{\partial E}{\partial y_j^{(p)}} = \frac{\partial E}{\partial z_i^{(p+1)}} \underbrace{\frac{\partial z_i^{(p+1)}}{\partial y_j^{(p)}}}_{w_{ij}^{(p+1)}} = w_{ij}^{(p+1)} f'(z_i^{(p+1)}) \frac{\partial E}{\partial y_i^{(p+1)}} \tag{10}$$

The calculation of the above term starts from the output layer for which:

$$\frac{\partial E}{\partial y_O} = 2y_O; \; f'(z_O) = 1 \tag{11}$$

and values for the rest of the layers are computed using backpropagation. Also note that some of the features in the hidden layers can be dropped (i.e., discarded) depending on the threshold of weights permitted, which is referred to as Dropout – a method used to prevent overfitting (i.e., "memorizing" the training data) [78].

## V. SENSITIVITY OF MACHINE LEARNED MODEL ON NETWORK AND TRAINING PARAMETERS

The effect of deep neural network hyperparameters, such as learning rate, activation function, size and depth of the network, batch size and dropout on trained model accuracy is investigated. There are several automatic hyperparameter tuning modules available online [75]. However, the authors opted for a manual parametric study to better understand how each of the parameters effect the training and accuracy of the model. To develop the most accurate network, ML models are trained in the interpolation mode, i.e., Hill dataset is included during training. The model accuracy is evaluated using *a priori* tests, i.e., the entire Hill DNS data is used to generate the input feature set and $P_K$ is queried from a trained ML model. The effectiveness of ML model training is monitored using evolution of training and validation residuals per epoch. The models are trained using up to 5000 epochs, but in most of the cases 3000 epochs are found to be sufficient, i.e., the training and validation residuals shows only a fractional drop in the last 1000 epochs with final residual values of around $10^{-8}$. Model accuracy is assessed qualitatively by comparing the ML model predicted $P_K$ contours with DNS data, and quantitatively by evaluating $R^2$ and slope of the correlation between models $P_K$ predictions with DNS data. Table 2 provides a summary of the parametric study along with key conclusions, and detailed discussions are provided below.

Learning rate has a significant effect on the training and accuracy as shown in Fig. 8. The largest learning rate of $10^{-2}$ shows a very fast initial convergence, and the residual drops to $5\times10^{-8}$ in 400 epochs. However, the trained model performs poorly (with $R^2 < 35\%$). The smallest learning rate ($10^{-3}$) performs better both in convergence and accuracy with $R^2 \sim 75\%$. However, the model strongly overfits the training data, evidenced by the validation residual being an order of magnitude larger than the training residual. The best predictions are obtained using a decaying learning rate, where the learning rate is reduced by an order of magnitude when the training residual does not change significantly for 100 consecutive epochs. The effects of activation function are also studied considering most commonly used linear (*ReLU*) and no-linear (*TanH* and *Sigmoid*) functions. Both non-linear activation functions perform equally well and were better than *ReLU*. Note that other variants of *ReLU* and non-linear functions are also available, but were not evaluated. This is because authors identified that other hyperparameters, as studied below, have stronger impacts on the ML training than choice of activation function.

Initially the model is trained using networks with different depths but with same total number of neurons, i.e., the product of # neurons and # layers remained the same, are compared. It is found that deeper networks (more hidden layers) work better than wider networks (more neurons per layer). The networks with increasing and then decreasing # neurons perform even better. Networks with five or more hidden layers are identified to be reasonable for this study and are further investigated. For this purpose, three network configurations are considered for further analysis, namely: (1) five layers with 96-128-256-128-96 neurons, (2) six layers with 96-128-256-256-128-96 neurons, and (3) seven layers with 96-128-256-512-256-128-96 neurons. As shown in Fig. 9, results improve significantly between five-layer and six-layer networks, and $R^2$ improves by 20%. The $R^2$ predictions do not show any significant improvements between six- and seven-layer networks, however, the correlation line slope is 10% better for the latter. It is observed that data batch size can help improve model accuracy. Note that the smaller batch size increases the training time since the network weights are modified more frequently. However, the objective here is to focus only on the model performance and not on computational expense. As shown in Fig. 10, using a smaller batch size results in better training (lower training residual), but the validation residual increases, resulting in an overfit model. A close inspection of production contour shows that the prediction of primary peak improves, but the prediction of "negative" production region degrades. The use of a smaller batch size is more effective for smaller networks (five-layer network), and training using a batch size of 2K results in $R^2$ of 0.8. Both six- and seven-layer networks show no significant improvements in predictions when the batch size is decreased from 15K to 5K, and both networks show very similar predictions using batch size of 5K. Thus, there is an optimal mini-batch size corresponding to the network size that provides best training and accuracy. The deterioration in the accuracy of the model for large mini-batch size is consistent with those reported in the literature [79]. However, increase in overfitting for smaller mini-batch size is a new observation. Herein, a network with six-layers and batch size of 5K is deemed optimal and is used for the remainder of this study.

Training of a ML model requires that all input and output variables be normalized, and for this model, a scale range of [-1,1] is used. This can be achieved by linearly scaling a variable ($V$) using the maximum ($V_{max}$) and minimum ($V_{min}$) values, and the scaled variable ($V_s$) is defined as:

$$V_s = 2\frac{V-V_{min}}{V_{max}-V_{min}} - 1 \qquad (12a)$$

Note that in this scaling, referred to as monolithic scaling, the mean feature $V_{mean} = (V_{max} + V_{min})/2$ corresponds to $V_s = 0$. For variables that have both positive and negative values, which is true for all variables except $Re_l$, a separate scaling can be used depending on its sign, i.e.,

$$V_s = \begin{cases} \frac{V}{V_{min}} & \text{for } V \leq 0 \\ \frac{V}{V_{max}} & \text{for } V > 0 \end{cases} \qquad (12b)$$

The variables can also be power scaled, which spreads the data more uniformly over the minimum and maximum range (Fig. 11, left column), before scaling between -1 and 1 using either the monolithic or separate approach. For this purpose, ($1/n$) power scaling is considered, where $n$ is an odd number. This allows retention of the sign of the variable. A combination of linear and power scaling is investigated to improve the accuracy of ML model. Analysis shows that a ML model trained using separate linear scaling of features show a 23% higher $R^2$ ~0.86 than models trained using monolithic linear scaling ($R^2$ ~0.63). It is found that power scaling all the variables degrade accuracy of the model to $R^2$ ~0.75. Power scaling of the input variables, but linear scaling of the output variable shows a significant deterioration of prediction ($R^2$ < 0.3). On the other hand, linear scaling of input features and power scaling of output features show somewhat better predictions ($R^2$ ~ 0.69), but the models are still less accurate than those obtained using linear scaling for both input and output features. Power scaling of $Re_l$ only provides improvements in the model accuracy. As shown in Fig. 11, the accuracy of the model improves when power scaling is varied from 1/3[rd] (for which $R^2$ ~0.84) to 1/9[th] (for which $R^2$ ~0.9), but 1/11[th] scaling does not show further improvement (for which $R^2$ ~0.89). Thus, the best scaling approach for our datasets is when all variables (except $Re_l$) are scaled separately in positive and negative regions, and $Re_l$ is first scaled using 1/9[th] power and subsequently monolithically scaled between [-1,1].

The effect of data clustering on model accuracy is also investigated. For this purpose, the training dataset is clustered using Python's scikit-learn module with four different clustering radii $r_c$ = 0.025, 0.05, 0.075 and 0.1. As shown in Fig. 12 (a-d), the clustering operation identifies different flow regimes (such as separation regime, acerating flow regime, boundary layer regime) across all flow datasets correctly. One observes that the first 200 clusters cover a major portion of the flow regime (Fig. 12b). In contrast, the small separation regimes (clusters 2400 to 2800, Fig. 12d) and the accelerating flow regimes (clusters 2800 to 3000. Fig. 12c) contain similar number of cluster samples. These results suggest that clustering can help in disambiguating and emphasizing the complex flow regimes well. As the clustering radius decreases, the number of clusters increase, i.e., 17346, 4190, 1591 and 800 clusters for $r_c = 0.025$, 0.05, 0.075 and 0.1, respectively. As summarized in Fig. 12(e), when 50% of data is used from all the clusters, trained models obtained using different $r_c$ provide similar model accuracy ($R^2 \sim 0.91$ to 0.92), but the correlation line slope decreases slowly from 0.92 to 0.88 with increasing $r_c$. However, when only 30% of the data from each cluster is used for training, the model accuracy decreases dramatically with increase in $r_c$. Overall findings indicate that smaller $r_c$ (resulting in larger number of clusters) are preferable, since a smaller percentage of data can be sampled from each cluster without losing accuracy. ML models trained using 1/3$^{rd}$ of cluster data obtained with $r_c \leq 0.05$ have the same accuracy ($R^2 \sim 0.92$) as models trained using the entire training dataset; thus, the clustered data approach is used hereafter.

Because ML model training is non-deterministic, repeated training using the same hyperparameters and data scaling results in stochastically different model accuracies. To quantify the repeatability uncertainty on ML model accuracy, five separate models are trained (Fig. 13). Correlation between models from different runs vary between 90 to 92%, and model accuracy shows a 4% variation over the repeated runs. In addition, as shown in Fig. 13 (left panel), the $P_K$ distribution predicted by ML model shows a distinct discontinuity around $X = 7$. Careful inspection of the input features reveals that the contours of vertical velocity direction ($\frac{\bar{v}}{|u|}$) has a similar discontinuity and is thus identified as the likely source of discontinuity. As demonstrated in Fig. 13 (right column), removing $\frac{\bar{v}}{|u|}$ as an input feature resolves the issue, and the averaged model accuracy is essentially the same as the models trained including $\frac{\bar{v}}{|u|}$. However, the model trained without $\frac{\bar{v}}{|u|}$ reduces the repeatability uncertainty to 2%, somewhat lower than the uncertainty in model trained with $\frac{\bar{v}}{|u|}$.

## VI. Feature Importance.

The importance of input features on the ML model is estimated in *a priori* tests using the best ML model developed in Section V. This analysis is performed using three approaches. Approach 1: *get_score_importances*, which estimates feature importance by measuring how ML model accuracy (score) decreases when a feature is absent [76]. This is achieved by computing the scores for 10 iterations with different random sets of input feature sets. Approach 2: *permutation_importance*, which computes feature importance by measuring how score decreases when a feature is randomly shuffled [77]. This is achieved by taking average over 30 permutations for each feature. Approach 3: SHAP analysis, which provides how each input feature contributes to the value of the output feature along with the prediction expected for the null set (i.e., without any input features), which are denoted by E[f(x)] [77].

The first two approaches require an estimation of score. Both mean squared error and $R^2$ error are considered to compute the score, and both provide very similar feature importance estimates, providing confidence in the methods. As shown in Fig. 14, both approaches show a dominant dependence on $D_{uv}$ followed by $W_{uv}$. The variables $Re_l$ and $\frac{\bar{v}}{|u|}$ are third most important, and $D_{uu}$ and $\frac{\bar{u}}{|u|}$ are found to be least important.

SHAP values at three cross-sections, X = 0.75, 4.2 and 6.5, are shown in Fig. 15(a), and the averaged SHAP value magnitudes over the entire domain (converted to %) are compared with the feature importance estimates of Fig. 14. The SHAP analysis provides feature importance qualitatively similar to the other two estimates, but in general is more consistent with *get_score_importances*. All the approaches show that $D_{uv}$ is the most dominant feature, but they show some differences in the importance of other features. Both SHAP and *get_score_importances* show somewhat lower importance of $W_{uv}$, thus, the importance of other features is higher, and both approaches show that $Re_l$, $\frac{\bar{u}}{|u|}$ and $\frac{\bar{v}}{|u|}$ are equally important.

To evaluate the input feature importance in different flow regions, the SHAP values are analyzed locally in Figs. 15(b-h). The E[f(x)] and contribution of individual input features are shown as waterfall plot, an ordered histogram, which when added together provides the output value. The histograms are ordered in the sequence the input features are used during training. The SHAP values in the peak production region of the flowfields (Fig. 15b) show that, $D_{uv}$ has more than 2 times larger contribution than other features. Both $D_{uu}$ and $Re_l$ are second largest contributors, and $\frac{\bar{u}}{|u|}$ has minimal contribution. In the mid-channel region of the flowfields (Fig. 15c), $W_{uv}$ and $D_{uv}$ are most important followed by $Re_l$. In the backflow regions (Fig. 15d), $\frac{\bar{u}}{|u|}$ plays the most important contribution, and $D_{uv}$, $W_{uv}$ and $Re_l$ show similar low contributions. In the reattachment regions (Fig. 15e), $\frac{\bar{v}}{|u|}$ plays the most important role, followed by $Re_l$, $\frac{\bar{u}}{|u|}$ and $D_{uv}$. In the accelerating flow regions close to the wall (Fig. 15f), the $D_{uu}$ contribution is higher relative to the other flow regions. In the attached boundary layer on the top wall (Fig. 15g), both $D_{uv}$ and $W_{uv}$ are important, whereas the former decreases the output value, but the latter increases the output value. Note that in this region wall-normal velocity is negligible, thus both $D_{uv}$ and $W_{uv}$ are expected to be same. In this region, $Re_l$ is the next most important feature. In the secondary peak regions (Fig. 15h), the $D_{uu}$ contribution is higher relative to other flow regions. Overall, the SHAP value predictions are consistent with the expected flow physics, in particular: (1) $D_{uv}$ is the key flow feature in the free-shear region, (2) $D_{uv}$ and $Re_l$ are most important in boundary layers, (3) all flow features are important in the backflow region, (4) wall-normal velocity is key in identifying the extent of the reattachment region, and (5) $D_{uu}$ is the most important feature in accelerating flow regions.

## VII. Neural Network Model Predictions in Interpolation and Extrapolation Modes

Accuracy of the ML model in interpolation and extrapolation mode is assessed in both *a priori* and *a posteriori* tests. For this purpose, ML models are trained by including different percentages of Hill datasets. Including the entire Hill dataset amounts to 15% of the total training data (ML$_{H15}$) and is referred to as *full interpolation* mode. ML models are also trained using lower percentages of the Hill data ranging from 9% (ML$_{H8}$), 5% (ML$_{H5}$), 3%, 1.5%, 0.75% and down to 0% (ML$_{H0}$). The model ML$_{H0}$ is trained without any information of the Hill case, thus the model is in *full extrapolation* mode. The rest of the cases are in *partial extrapolation* modes that are in between the *full interpolation* and *full extrapolation* modes.

### A. A priori Tests

For the *a priori* tests, ML models are trained including various percentages of the Hill dataset. Then, DNS data of the entire Hill domain are used as the input feature set ($\frac{\bar{u}}{|u|}, \frac{\bar{v}}{|u|}, D_{uu}, D_{uv}, W_{uv}$ and $Re_l$) and used to query for ML-learned $P_K$ over the entire Hill flow field using the ML models including those trained without $\frac{\bar{v}}{|u|}$ as an input feature. The accuracies of the ML model $P_K$ predictions are assessed via correlations to the DNS data (Fig. 16a, left panel and Fig.

16b), and by qualitatively comparing $P_K$ contour plots (Fig. 16b, right panel) to the DNS contours (Fig. 4). Plots of ML model training and validation residuals are shown in Fig. 16a (middle panel).

The *full interpolation* model $ML_{H15}$ predicts the peak $P_K$ in the shear layer, the secondary peak upstream of the uphill and negative $P_K$ at the base of the uphill region very well, and predictions correlate within 87% of the DNS. In contrast, the *full extrapolation* model $ML_{H0}$ performs poorly, except in the upper wall boundary layer region, and has less than 5% correlation with DNS data. ML model accuracy improves sharply as even a small percentage of Hill data is included, i.e., $ML_{H5}$ shows around 80% correlation with DNS data. The $ML_{H9}$ model shows a somewhat better $R^2 = 83\%$ than $ML_{H5}$, but the averaged correlation (obtained by averaging $R^2$ and correlation line slope) is same as $ML_{H5}$. As shown in Fig. 16(b), the *partial extrapolation* models trained without $\frac{\bar{v}}{|u|}$ overall show 10% lower correlation compared to the models trained with $\frac{\bar{v}}{|u|}$. However, both the models show similar accuracy in the *full interpolation* model.

Overall, the *a priori* test demonstrates that ML model performs poorly in *full extrapolation* mode, which is expected, however, just incorporating a nominal 5% of the test flow datasets (selected based on data clustering) significantly enhances the ML model accuracy. No definite conclusions can be drawn regarding the effect of $\frac{\bar{v}}{|u|}$ as an input feature on the ML model accuracy. On one hand, the ML models trained with $\frac{\bar{v}}{|u|}$ are generally better than those without $\frac{\bar{v}}{|u|}$, however, for the *full interpolation* mode, the latter is somewhat qualitatively better than the former.

## B. A posteriori Tests

For the *a posteriori* tests are performed using medium grid, wherein $ML_{H0}$ to $ML_{H15}$ models are used to predict $P_K$ fields each CFD time step and used in the solution of the $k$ and $\omega$ equations (shown in Eqs. 2b). To access the accuracies of the ML models, the predictions of the streamwise velocity and flow streamline, TKE and $P_K$ are compared with DNS data in Fig. 17(a), and the correlations between ML model predictions and DNS data are summarized in Fig. 17(b). As the percentage of Hill data is reduced during training, the extent of the flow separation region is overpredicted compared to DNS, and the correlations for the TKE and $P_K$ predictions decrease. Both $ML_{H15}$ and $ML_{H9}$ models show correlations of 97%, 70% and 40% for streamwise velocity, TKE and $P_K$, respectively. $ML_{H5}$ shows somewhat lower correlations of 95%, 64% and 30%. The RANS predictions show averaged correlations of 97%, 64% and 35% for streamwise velocity, TKE and $P_K$, respectively. Thus, both ML$_{H15}$ and ML$_{H9}$ show better predictions than RANS, and ML$_{H5}$ predictions are comparable to those of RANS. In general, ML models that show less than 50% correlations with DNS data in the *a priori* tests result in larger separation bubbles and less accuracy than RANS in the *a posteriori* tests.

Figure 18 compares Y profiles of mean streamwise velocity, TKE, shear stress and $P_K$ at several X cross-sections using data from DNS, RANS, and ML$_{H5}$ with and without $\frac{\bar{v}}{|u|}$. The results indicate that the ML models perform somewhat better than RANS for the velocity profiles especially in the reattachment region (X = 6). However, the ML models show significantly better TKE and shear stress predictions than RANS. The improvements from the ML model can be directly attributed to the enhanced TKE production over those predicted in RANS. In addition, the ML models predict the negative $P_K$ in the accelerating flow region at the foot of the uphill (X=8), which not predicted in RANS. Further note that the ML model trained with $\frac{\bar{v}}{|u|}$ slightly outperforms the model trained without $\frac{\bar{v}}{|u|}$ for TKE and shear stress predictions. However, note that in the *a priori* tests, ML models trained with and without $\frac{\bar{v}}{|u|}$ show similar advantages and disadvantages. Thus, no concrete conclusions can be drawn regarding the role of $\frac{\bar{v}}{|u|}$ on ML model accuracy for these datasets.

## VIII. SUMMARY, CONCLUSIONS, AND FUTURE WORK

This objective of this study is to evaluate the predictive capability of machine-learned turbulence models in extrapolation modes. For this purpose, flow over a periodic hill is used as a test case, which involves complex flow regimes such as attached boundary layer, free-shear layer, flow separation and attachment, and accelerating flow. For this test case, the commonly used *k-ω* SST RANS model predicts the mean velocity reasonably well, although the extent of the separation length is overpredicted. The biggest limitation of the RANS model pertains to the prediction of TKE and shear stress, which are underpredicted by 65% compared to DNS data. Deficiencies of RANS modeling is due to significantly lower than realistic TKE production. To address this limitation, machine learned models are trained to predict turbulence production and then coupled with a CFD solver. The predictive capability of the ML model is tested for both *a priori* and *a posteriori* tests.

To train machine learned models, DNS/LES datasets are curated for eight different cases with flow separation and attachment regimes, and seven of the datasets have a much smaller separation regions and lower turbulence compared to the Hill case. This study investigates how accuracy in *a priori* tests and predictability in *a posteriori* tests change as various percentages of the Hill DNS dataset are included during the training of ML models, ranging from 0% (*full extrapolation* mode) to 100% of the Hill dataset (*full interpolation* mode). The study also performs a parametric study to investigate several aspects of ML model training, including (1) the effect of training and neural network hyperparameters such as learning rate, network size, activation function, mini-batch size, on training/validation residual and model accuracy; (2) the quantification of the uncertainty in model accuracy due to the non-deterministic nature of neural network training; (3) the effect of data scaling and data clustering on ML model accuracy; and (4) feature importance to quantify correlations between the input features and the turbulence production in different flow regimes.

Several conclusions are made for turbulence enhancement using ML modeling for the flow datasets investigated. The parametric study of hyperparameters indicates that using an adaptive learning rate significantly improves ML model training and accuracy compared to a fixed learning rate. Non-linear activation functions such as *TanH* and *Sigmoid* perform much better than the linear *ReLU* function. Deeper networks are much better than the wider networks. However, networks with increasing then decreasing #neurons per layer are the best. The mini-batch size that generates the most accurate ML model is strongly related to the network size. In general, larger batch sizes work well for larger networks, and smaller batch sizes work well for smaller networks. However, smaller than optimal batch sizes result in overfitting, whereas larger than optimal batch sizes undermine accuracy. Data scaling significantly affects model accuracy. Scaling of features differently for negative and positive values of data distributions is more effective that monolithic scaling using maximum and minimum values. Input features that have highly skewed distributions between its minimum and maximum can be scaled using power laws to improve (widen) their distribution and improve the accuracy of ML models. Data clustering is found to be an efficient approach to avoid bias of a model towards more prevalent flow regimes, and in general enables more efficient training since much redundant and voluminous data can be clustered together. Neural network training is non-deterministic and can result in up to 4% uncertainty in model accuracy. Feature importance analysis identifies that: (1) $D_{uv}$ is the key feature that generates high $P_K$ in free-shear regions, (2) both $D_{uv}$ and $Re_l$ are important in boundary layers, (3) velocity direction plays a key role in backflow regions, (4) wall-normal velocity is key in identifying the extent of reattachment regions, and (5) $D_{uu}$ is the most important feature in accelerating flow regimes.

Several conclusions are made about ML model performance in *a priori* and *a posteriori* testing. ML models performed poorly (< 10% correlation with DNS) in full extrapolation mode. ML model predictions improved significantly in partial extrapolation mode as percentages of the Hill data was added during training. Addition of a minimal amount of Hill data, namely, 5% of its dataset, results in a ML model with an 80% correlation to DNS data. ML models trained using the entire Hill data (*full interpolation* mode) has 87% correlations to DNS data. It is found that ML models that have 80% or more accuracy in *a priori* testing provide good predictive accuracy in *a posteriori* testing, and such models perform better than RANS models for the TKE and shear stress predictions.

Overall, ML modeling for TKE production is deemed to be a reliable approach to enhance the predictive capability of RANS models. ML models do not result in numerical instabilities even when the models have poor accuracy in *a priori* testing. In addition, ML models allow for the prediction of negative turbulence production in accelerating flow regimes, which is not modeled in linear RANS models. In general, it is recommended that before a ML model is used for *a posteriori* application in CFD codes, (1) quartile analysis of the input and output features should be performed for the training and test datasets to ensure that the ML model will not be operating in a *full extrapolation* mode, and (2) *a priori* tests should be performed (using RANS solutions or DNS datasets) to ensure that the ML model has reasonable accuracy. A review of the literature indicates that ML models have not been applied for unsteady flows, which impose additional complexity and expenses as the turbulence regime evolves constantly. Future work will focus on such flows, such as unsteady high Re flow over a cylinder.

**SUPPLEMENTARY MATERIAL**

Supplementry matrial is provided to aid the reviewer and is not for publications. The paper provides only the key plots or the summary plots, which are self sustained for publication. The supplement provides additional figures and tables that helped in drawing conclusions.

**ACKNOWLEDGMENTS**

**Table 1**: DNS/LES database for separated flow used in this study to train a ML model for turbulence production.

| | Dataset | Flow Property | #Datapoint | Raw Variables | Input features | Output features | Notes |
|---|---|---|---|---|---|---|---|
| DNS | Hill | $U_b = 1$ m/s<br>$h = 1$<br>$Re = 10595$<br>$\nu = 2.415\text{e-}5$ m²/s | 303,264 | $x, y, \bar{u}, \bar{v}, \bar{p}$<br>$\overline{u'u'}, \overline{u'v'}, \overline{v'v'}$<br>$\overline{w'w'}, k, \nu_T,$ | | | 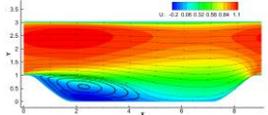 |
| | Hump | $U_c = 1$ m/s<br>$h = 0.128$<br>$Re_c = 936,000$<br>$\nu = 8.3467\text{e-}6$ | 280,918 | $x, y, \bar{u}, \bar{v}, \bar{p}$<br>$\overline{u'u'}, \overline{u'v'}, \overline{v'v'}$<br>$\overline{w'w'}$ | | | 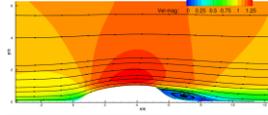 |
| | Nozzle | $U_c = 1$ m/s, $h = 2/3$<br>$h_c = 1$ (half channel)<br>$Re = U_c h_c/\nu = 12600$<br>$\nu = 7.93651$ m²/s | 887,041 | $x, y, \bar{u}, \bar{v}, \overline{u'u'}, \overline{u'v'}, \overline{v'v'},$<br>$\overline{w'w'}, \frac{\partial u}{\partial x}, \frac{\partial u}{\partial y}, \frac{\partial v}{\partial x}, \frac{\partial v}{\partial y}$ | | | 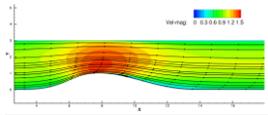 |
| LES | Curved backward facing step | $U_c = 1$ m/s<br>$h = 1$<br>$Re = 13700$<br>$\nu = 7.3\text{e-}5$ | 122,880 | $x, y, \bar{u}, \bar{v}, \bar{p}$<br>$\overline{u'u'}, \overline{u'v'}, \overline{v'v'}$<br>$\overline{w'w'}, k$ | $\frac{\bar{u}}{|u|}, \frac{\bar{v}}{|u|},$<br>$D_{uu}, D_{uv},$<br>$Re_l, W_{uv}$ | $P_k = -\tau_{ij} D_{ij}$ | 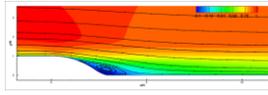 |
| | Bump, h20<br>$h = 0.02$ m<br>$c = 0.305$ m | | 122,880 | | | | 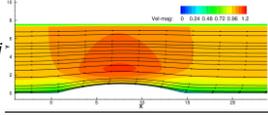 |
| | Bump, h26<br>$h = 0.0268$ m<br>$c = 0.305$ m | | 122,880 | | | | 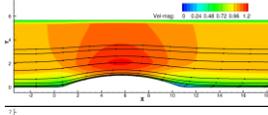 |
| | Bump, h31<br>$h = 0.0315$ m<br>$c = 0.305$ m | $U_{ref} = 16.77$ m/s<br>Inlet $Re_\theta = 2500$<br>$\theta = 0.0036$ m<br>$\nu = 2.415\text{e-}5$ m²/s | 122,880 | $x, y, \bar{u}, \bar{v}, \bar{p}$<br>$\overline{u'u'}, \overline{u'v'}, \overline{v'v'}$<br>$\overline{w'w'}$ | | | 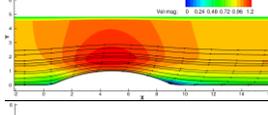 |
| | Bump, h38<br>$h = 0.0384$ m<br>$c = 0.305$ m | | 122,880 | | | | 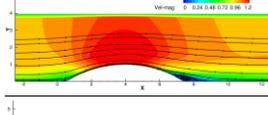 |
| | Bump, h42<br>$h = 0.042$ m<br>$c = 0.305$ m | | 122,880 | | | | 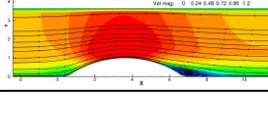 |
| *Total* | | | 2,208,503 | | | | |

**Table 2**: Parametric study to analyze the sensitivity of the machine learned models on neural network hyperparameters, data scaling and clustering.

| Parameter | | Key conclusion |
|---|---|---|
| Learning rate | $10^{-2}$, $10^{-2}$, Adaptive | • Adaptive learning rate is better |
| Activation function | *ReLU*, *TanH*, Sigmoid | • Both *TanH* and Sigmoid perform better than *ReLU* |
| # Hidden Layers and neurons | 5: 96-128-256-128-96<br>6: 96-128-256-256-128-96<br>7: 96-128-256-512-256-128-96 | • Deeper network works better than wider network<br>• Network with increasing and then decreasing # neurons is best |
| Min-batch Size | 1K, 2K, 5K, 10K, 15K | • Smaller than optimal mini-batch size result in overfitting, whereas larger than optimal mini-batch size reduces accuracy.<br>• Mini-batch size of 5K is optimal for 6-layer network. |
| Data scaling | Linear and Power scaling | • Separate scaling on +ve and -ve sides is more accurate than monolithic scaling using Maximum and Minimum.<br>• Power scaling of $Re_l$ improved accuracy. Best accuracy obtained for 1/9$^{th}$ power. |
| Data clustering | Cluster radius: 0.025, 0.05, 0.075, 0.1 | • 1/3$^{rd}$ of data for each cluster when $r_c \leq 0.05$ shows same accuracy as the model trained using the entire dataset. |
| Repeatability | Input features w/ $\frac{\bar{v}}{|u|}$ and w/o $\frac{\bar{v}}{|u|}$ | • Model accuracy shows 4% variation over repeated runs.<br>• Model trained w/o $\frac{\bar{v}}{|u|}$ resolves $P_K$ discontinuity issue, and reduces repeatability uncertainty. |

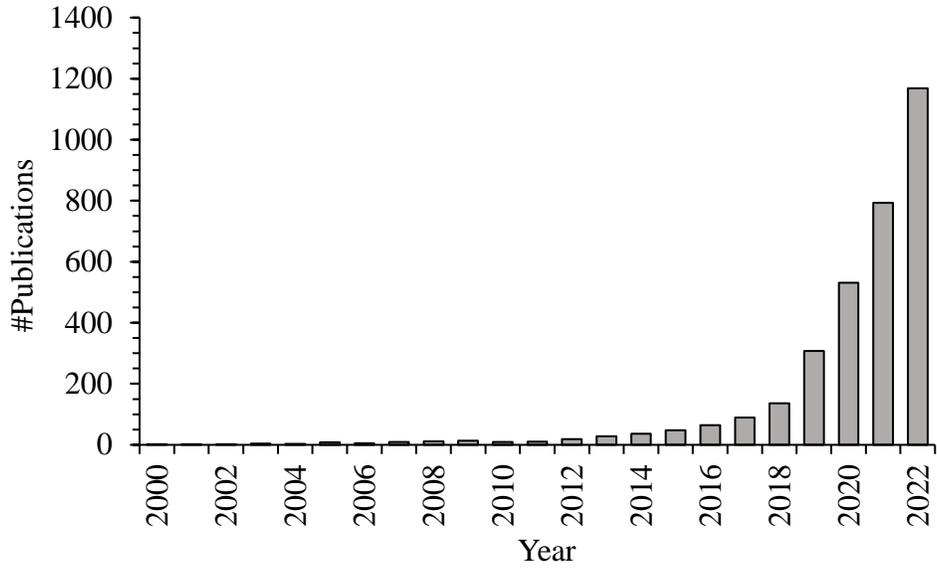

(a)

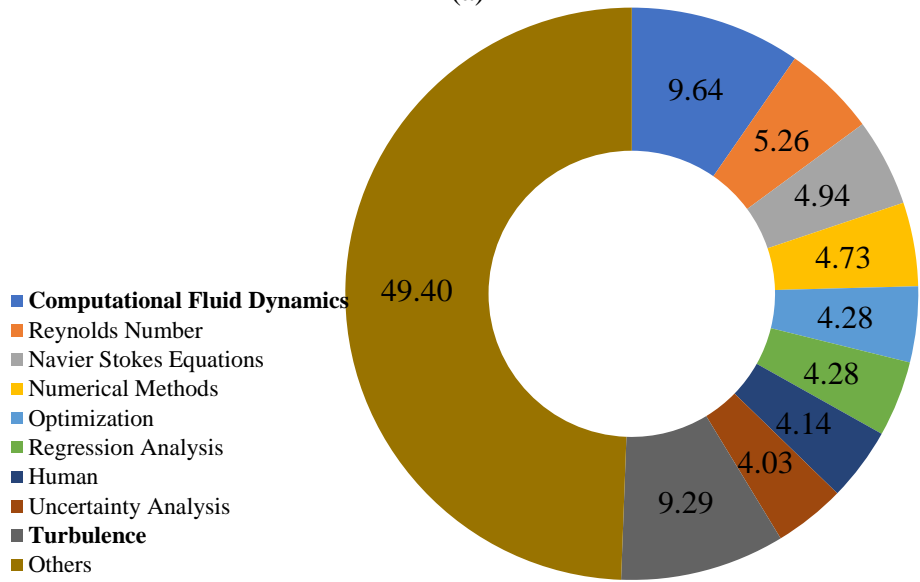

(b)

**Figure 1**: (a) Number of peer-reviewed publications with keyword "machine learning" and "fluid mechanics" in engineering in the past two decades. (b) The top 10 keywords in the publications. The data is obtained from scopus.com [1] in January 2023, and the keyword search is from publication title, abstract and keywords.

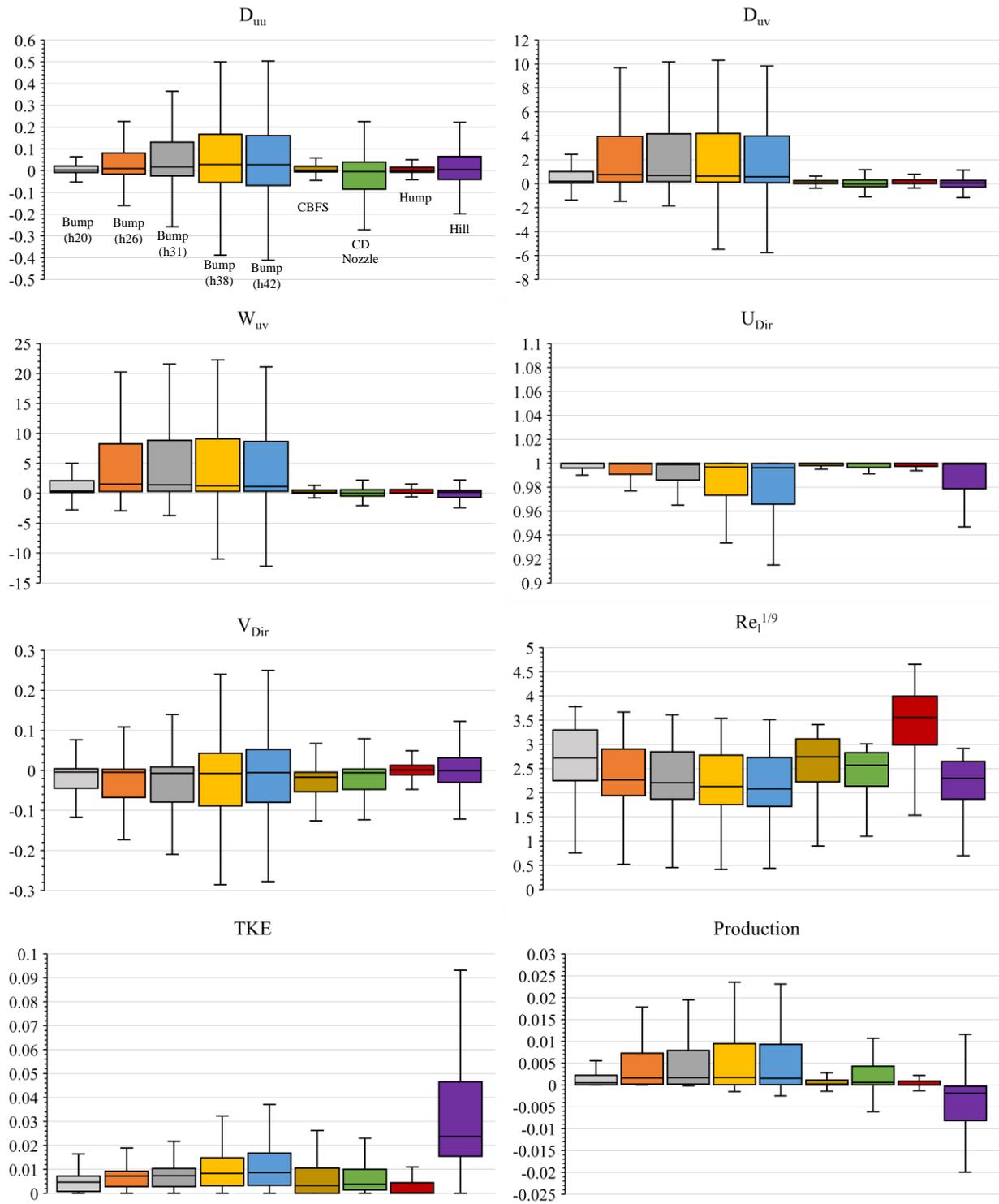

**Figure 2**: Box and whisker plot showing the data distribution through the quartiles. The shaded box region shows the data in 2nd and 3rd quartile, i.e., 50% of the data. The error bar shows the median of the 1st and 4th quartile of the data.

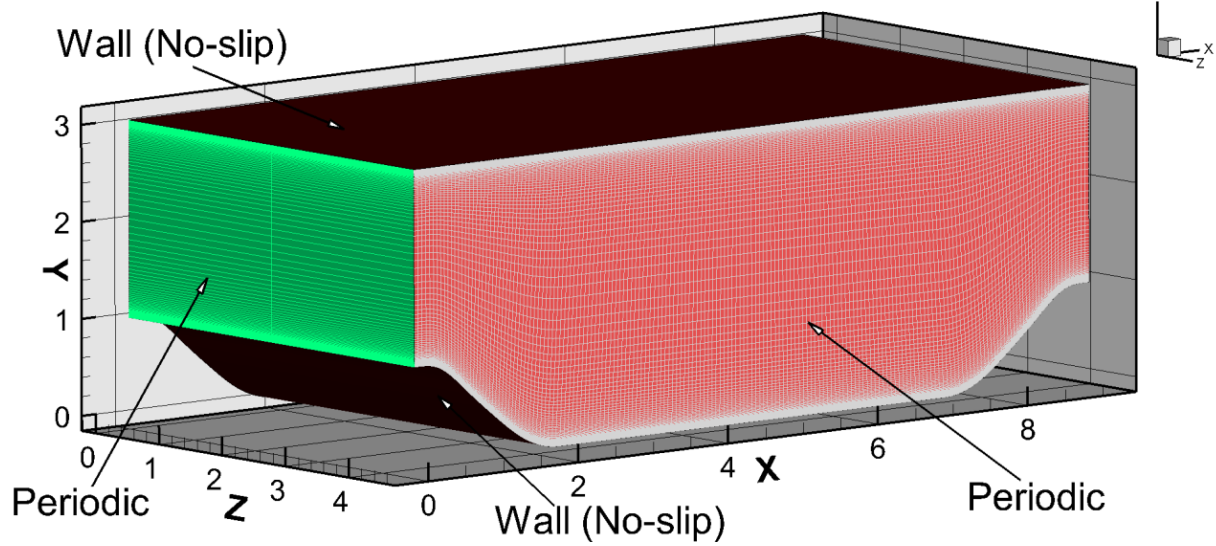

**Figure 3**: Domain, grid and boundary condition used for the periodic hill case. The spanwise (Z) direction is discretized using 2 cells to obtain a 2D flow.

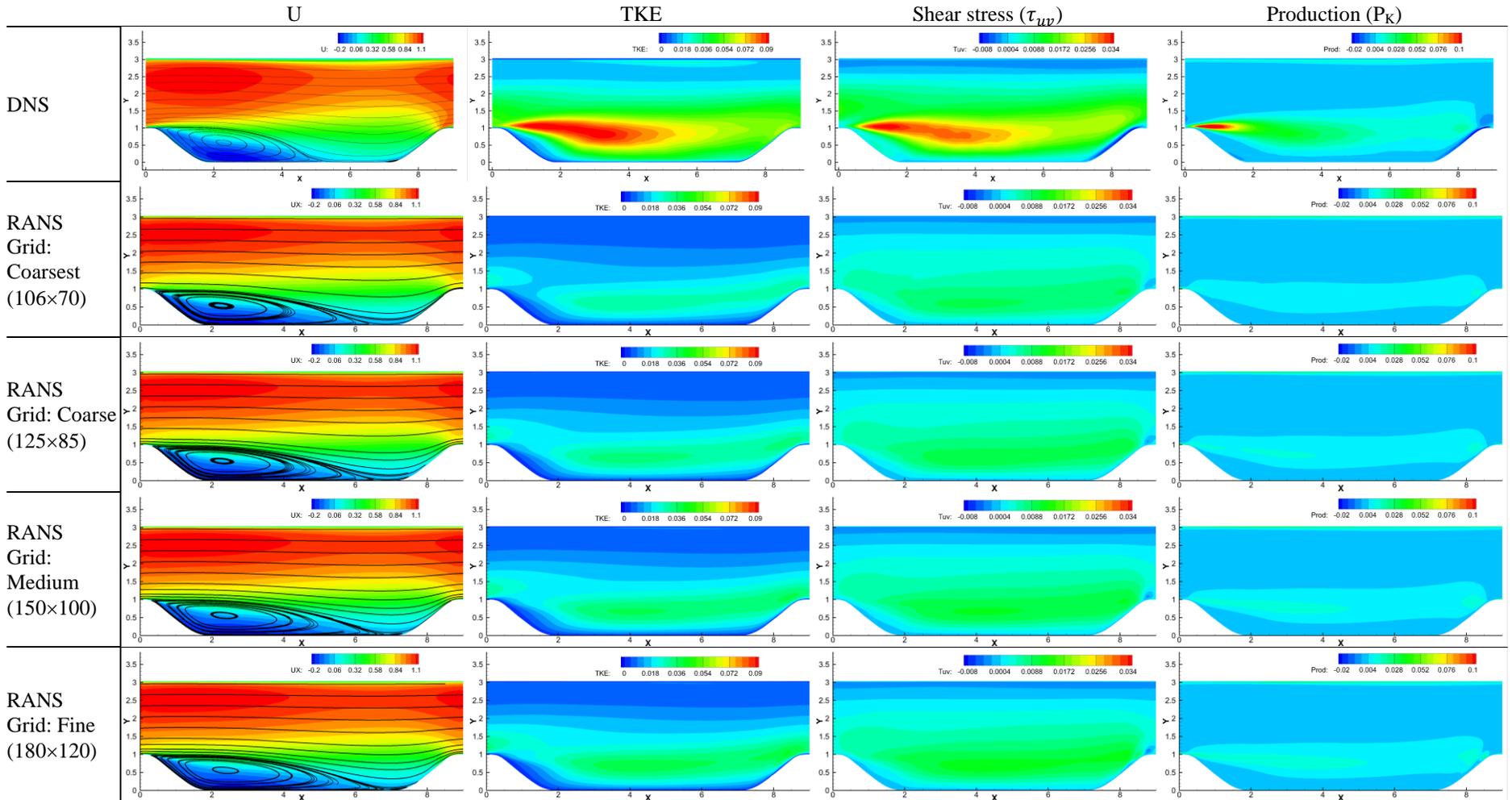

**Figure 4**: Comparison of contour plot of streamwise velocity (left column), TKE (2nd column from left), shear stress $\tau_{uv}$ (3rd column from left) and $P_K$ (right column) predicted by DNS and *OpenFOAM* RANS on Coarsest, Coarse, Medium and Fine grids.

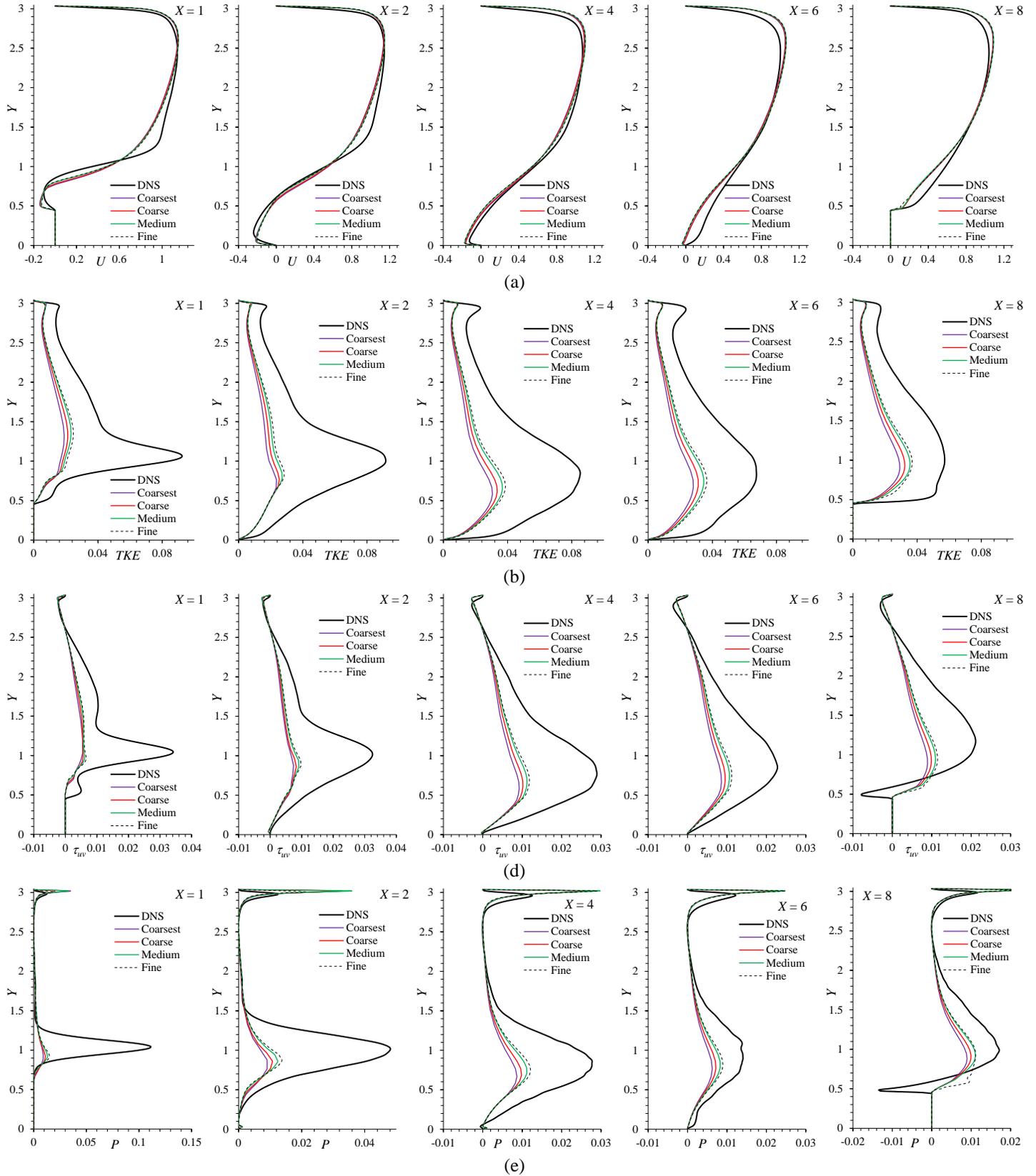

**Figure 5**: Line plots of (a) streamwise velocity, (b) TKE, (c) shear stress $\tau_{uv}$ and (d) $P_K$ predicted *OpenFOAM* RANS at several streamwise locations (X = 1, 2, 4, 6 and 8) on Coarsest, Coarse, Medium and Fine grids are compared with DNS results.

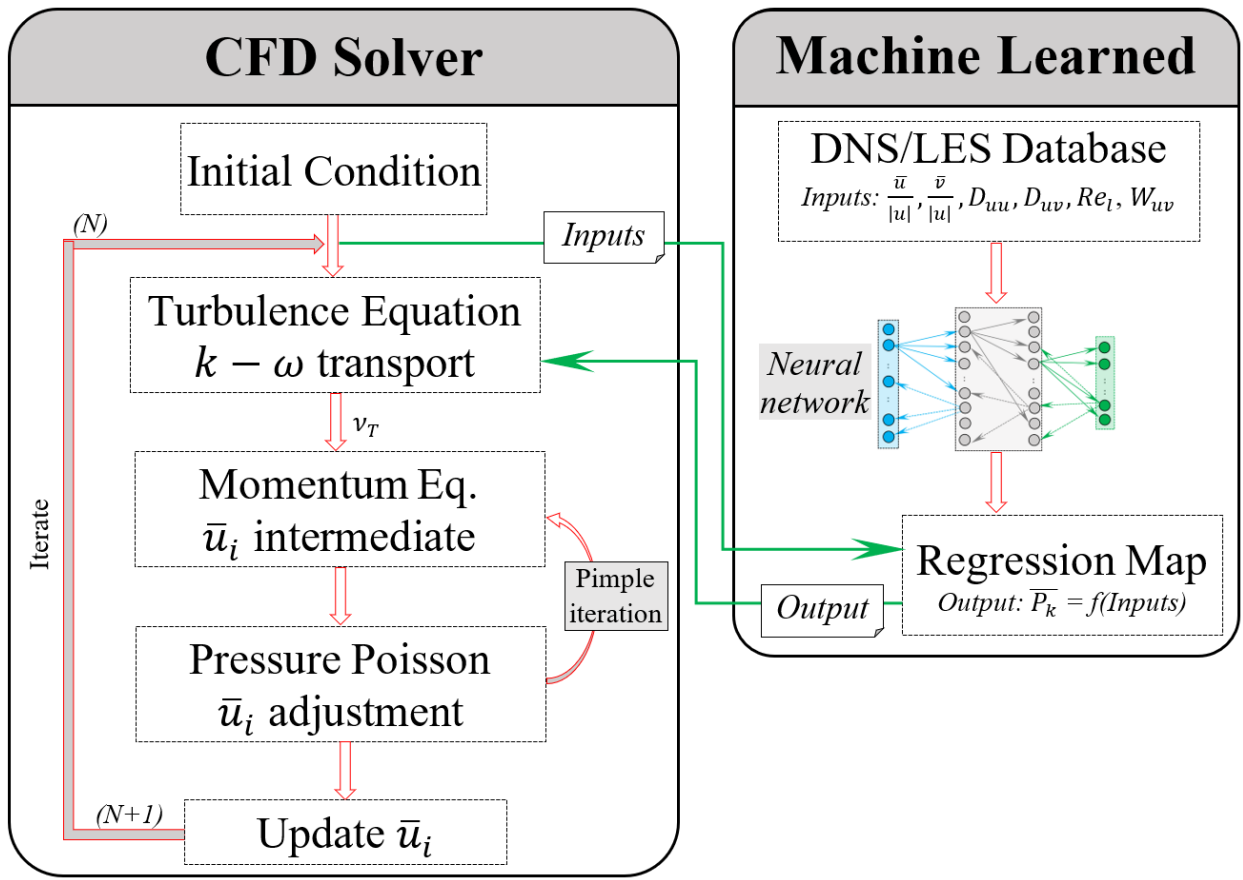

**Figure 6**: Schematic diagram depicting the RANS solution process and coupling with machine learned turbulence model. The neural network trained regression map is developed as a pre-processing step. The CFD solver is coupled with the neural network trained regression map for the turbulence production (output feature). The runtime query involves CFD solver providing the input features to the regression map, which in return provides the turbulence production to be used in the turbulence transport equations.

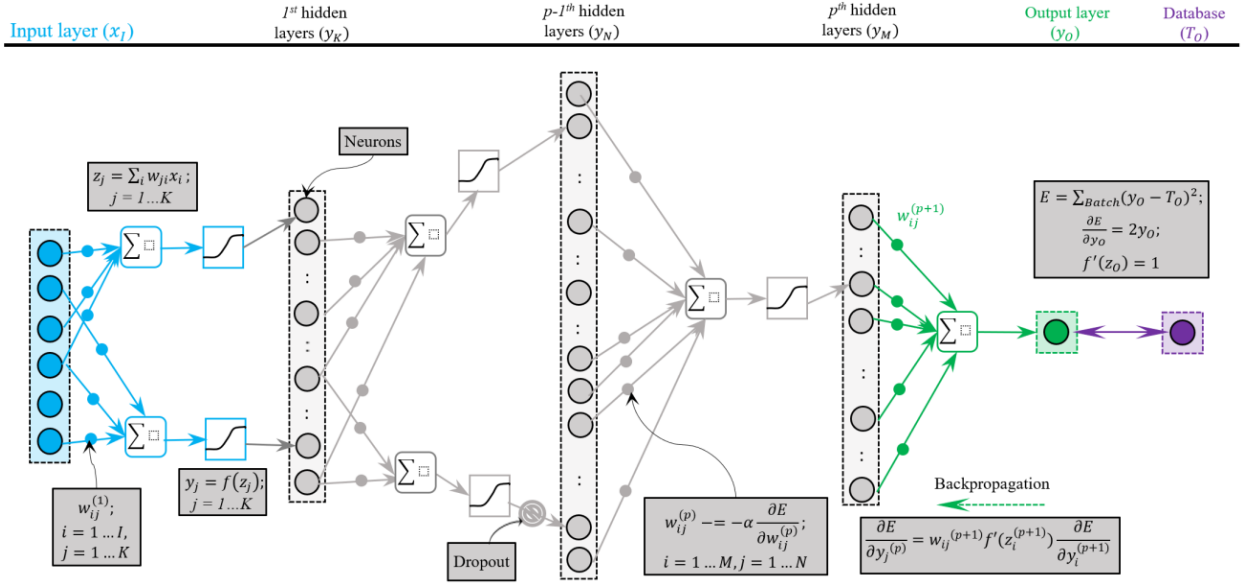

**Figure 7:** Block diagram summarizing the key aspects of neural network. Diagram is inspired from LeCun et al. [74]. The above example shows a neural network with one input layer (with 6 input features, $x_i$ $i= 1...I$), three hidden layers (for which #neurons in the layers first increase then decrease) and output layer with one feature ($y_O$). Each of the large circles represent a neuron or $K$ features ($y_i$, $i= 1...K$) in the hidden layer. The features in the hidden layers are obtained by weighted linear combination of the features in the previous layers and passing them through an activation function $f(z_j)$. Note that activation function is not used for the output layer. The small dots on each line represent the weight matrix ($w_{ij}$). The weights are adjusted via backpropagation of the network prediction error ($E$) computed using a predefined cost function, which compares the network output ($y_O$) with the true values ($T_O$) of the data batch. The derivatives of the errors are computed to adjust the weight matrix, where α is learning rate. Dropout is used to randomly remove specified percentage of neurons in each hidden layer during training.

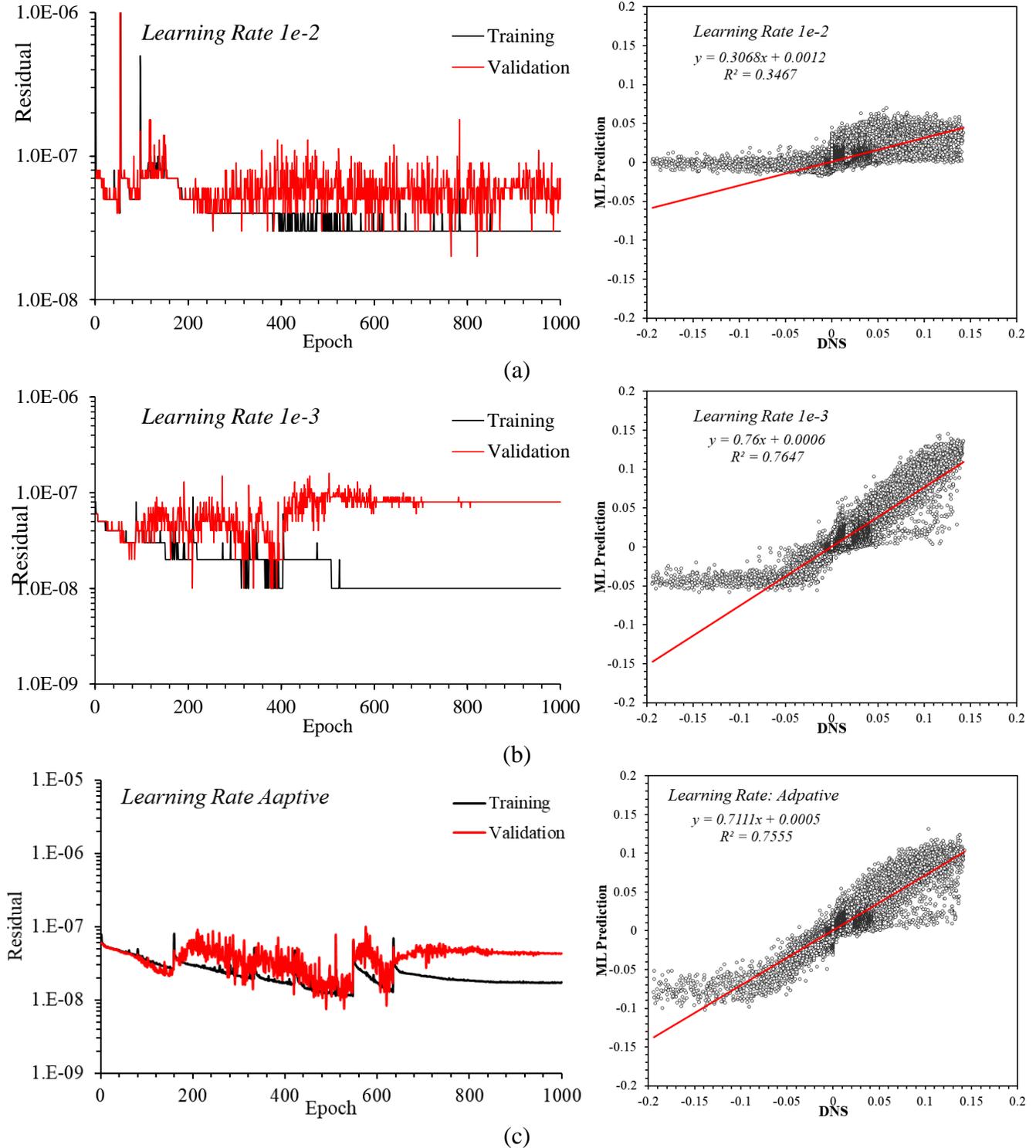

**Figure 8:** Parametric study to evaluate the effect of learning rate on model training and accuracy. Models were trained using all the 9 datasets using a neural network with 6 hidden layers, mini-batch size of 10K for 1K Epochs. $Re_l$ was scaled using $1/9^{th}$ power, and rest of the variables were linearly scaled.

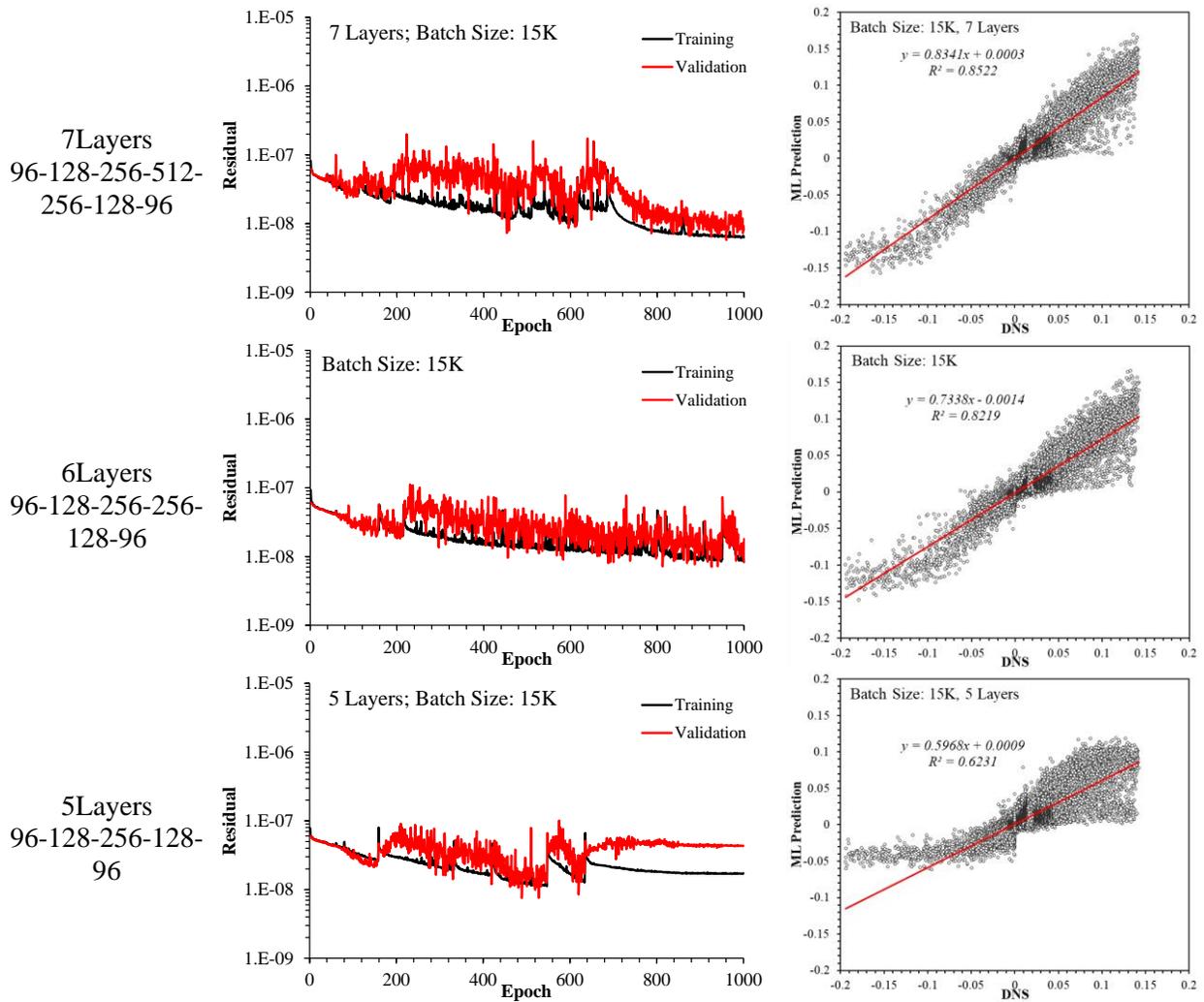

**Figure 9**: Parametric study to analyze the effect of network depth and size on ML training and accuracy. Models were trained using Tanh Activation function, 7-, 6- and 5-layer deep neural network and min-batch size of 15K, and 20% dropout between layers.

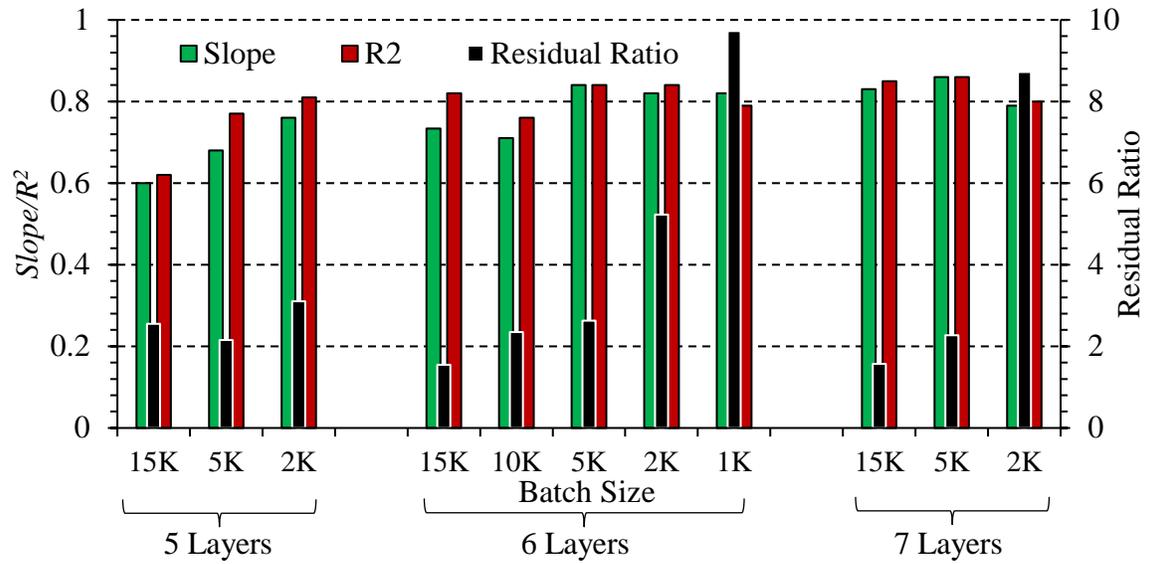

**Figure 10**: Summary of the effect of mini-batch size and number of layers on model accuracy. Models were trained using Tanh Activation function, 7-, 6- and 5-layer deep neural network and mini-batch size of 1K to 15K, and 20% dropout between layers.

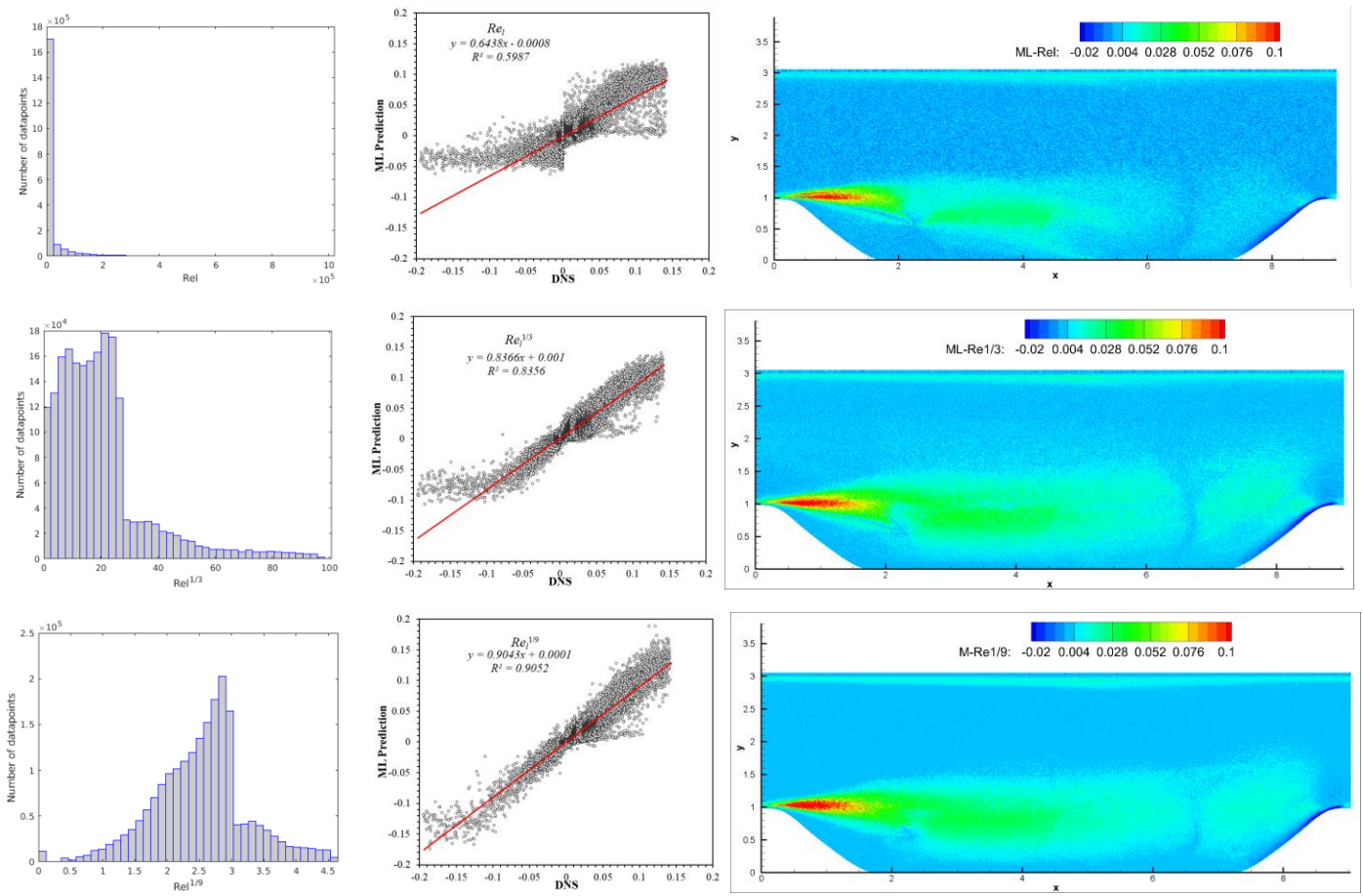

**Figure 11**: Effect of $Re_l$ scaling on model accuracy. Training was performed using Tanh Activation function, 6-layer deep neural network, mini-batch size of 5K, and 20% dropout between layers.

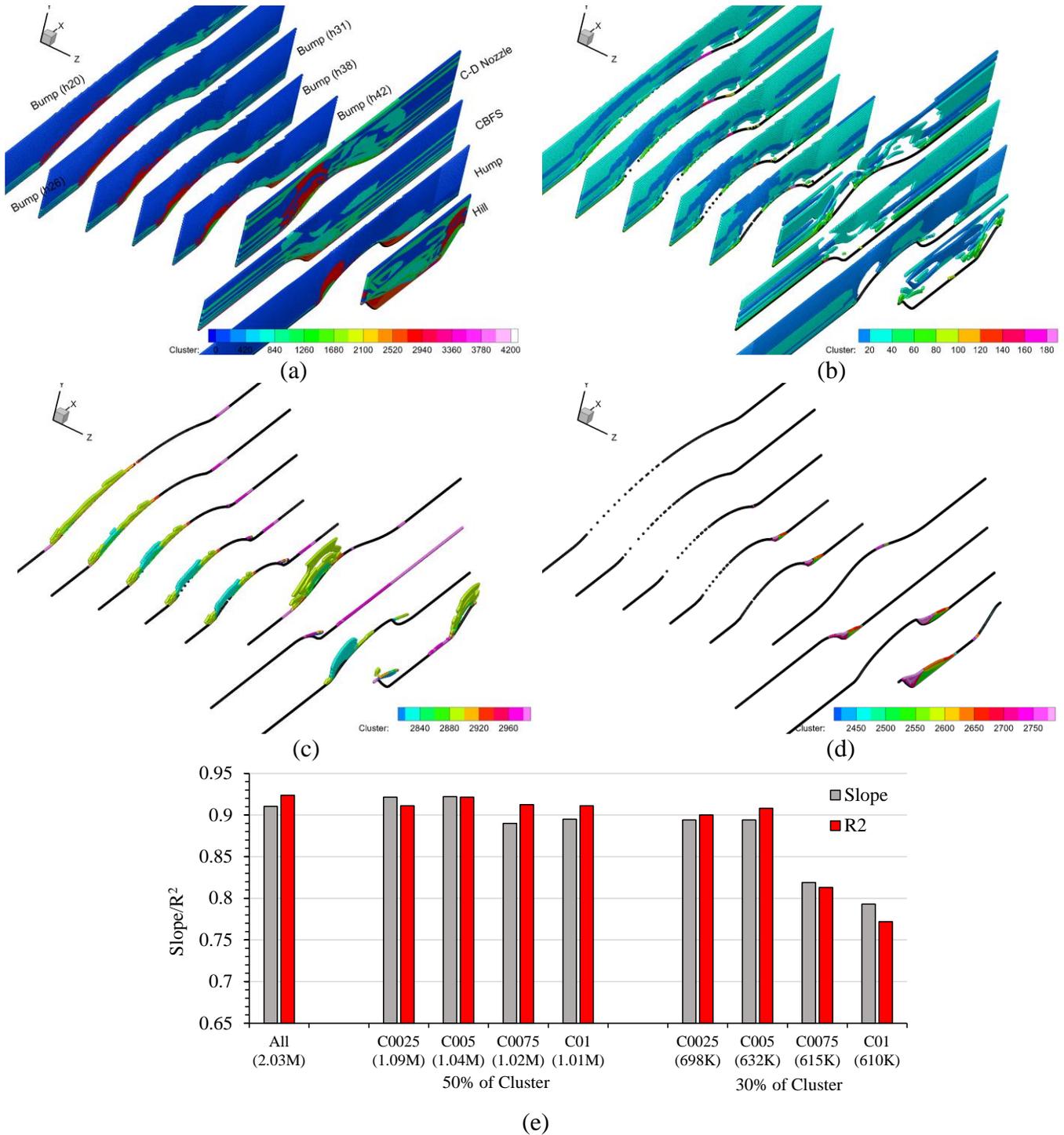

**Figure 12**: (a-d) Analysis of the clusters obtained for all the cases for $r_c = 0.05$. (a) Map of all the 4200 clusters; (b) map of first 200 clusters; (2) map of 200 clusters between 2800 to 3000 (accelerating flow regime); and (d) map of clusters between 2400 and 2800 (flow separation regime). (e) Summary of effect of data clustering and % of data used from each cluster during training on model accuracy.

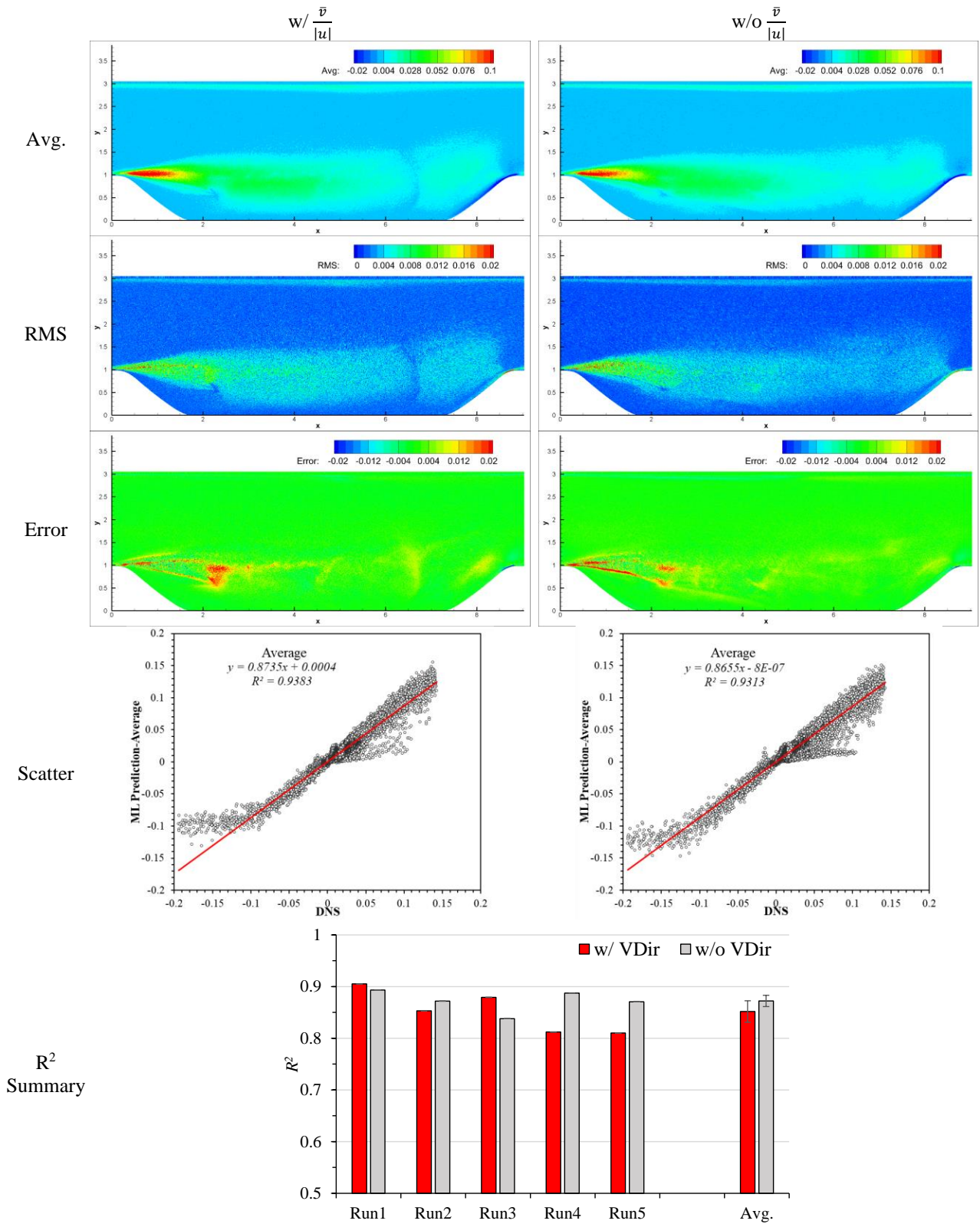

**Figure 13**: Assessment of training repeatability uncertainty on ML model accuracy. The uncertainty is quantified using five repeated runs with input features including $\frac{\bar{v}}{|u|}$ (left column) and w/o $\frac{\bar{v}}{|u|}$ (right column). The bottom plot shows summary of ML model correlation with DNS during repeated run.

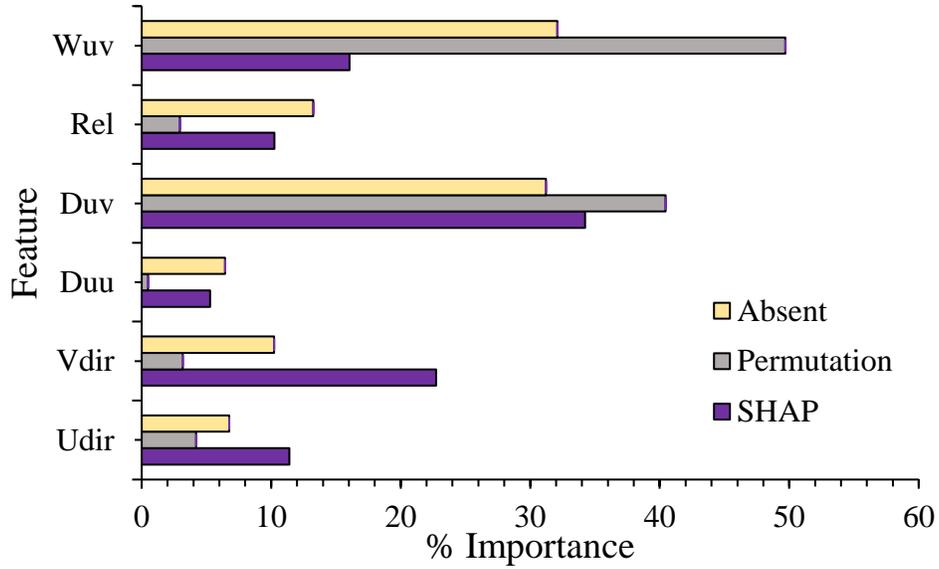

**Figure 14**: Feature importance estimates over the entire flow domain using: (a) *get_score_importances*, which measures the decrease in ML model accuracy (score) when a feature is absent (referred to as Absent). (2) *permutation_importance*, which measures decrease in score when a feature is randomly shuffled (referred to as Permutation); and (3) SHAP analysis, which measures the impact of a feature on score taking into account the interaction with other features. For the latter, the plot shows the % magnitude of the SHAP values for each feature.

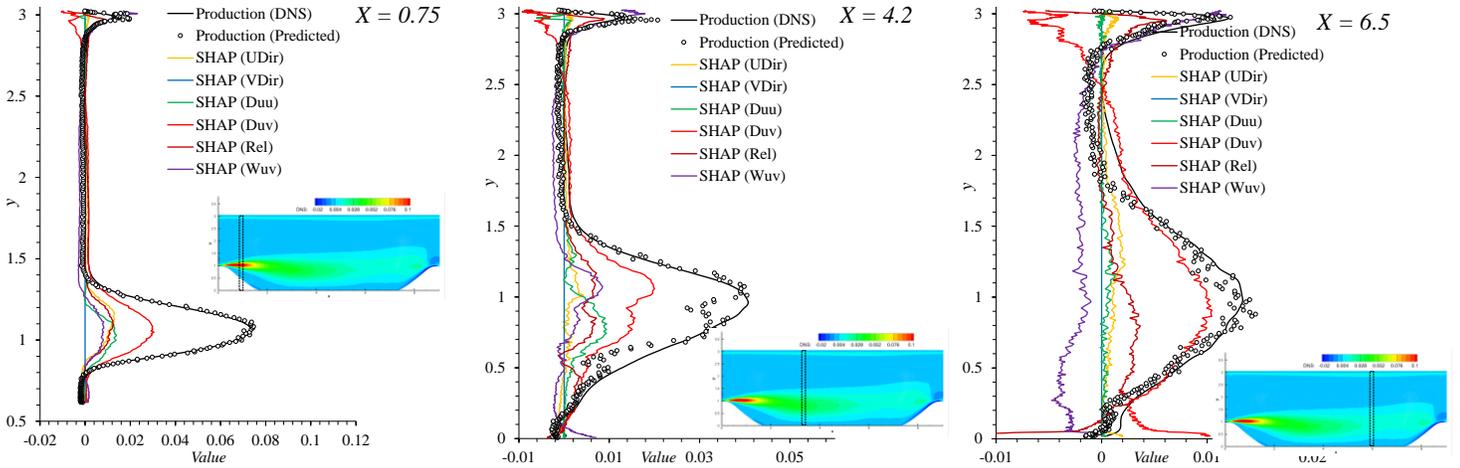

(a)

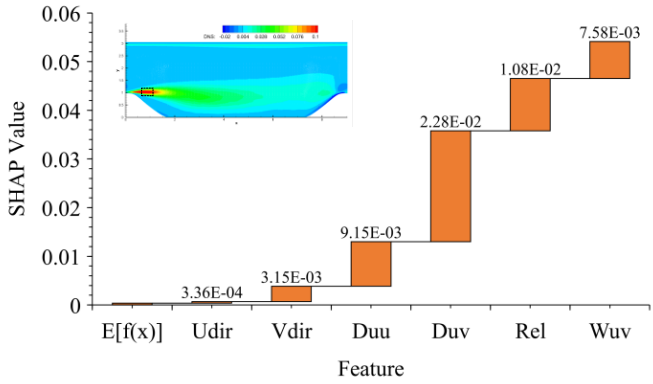

(b)

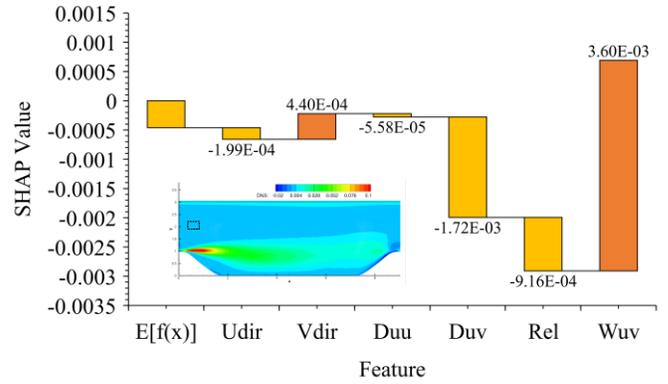

(c)

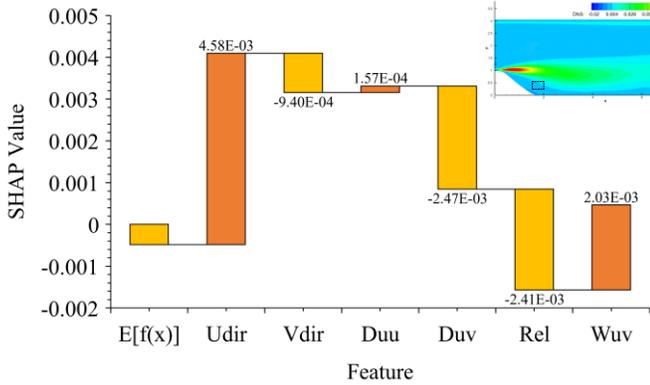

(d)

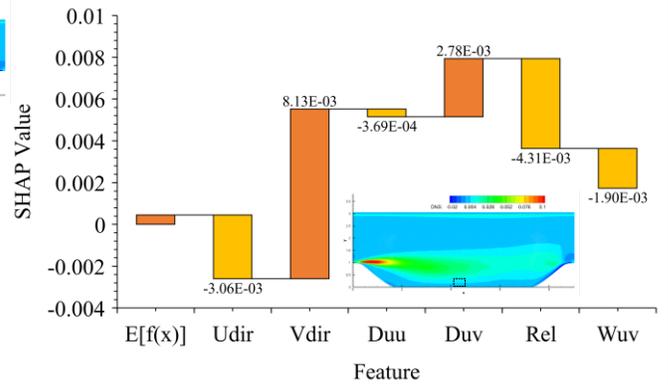

(e)

**Figure 15**: Continued.

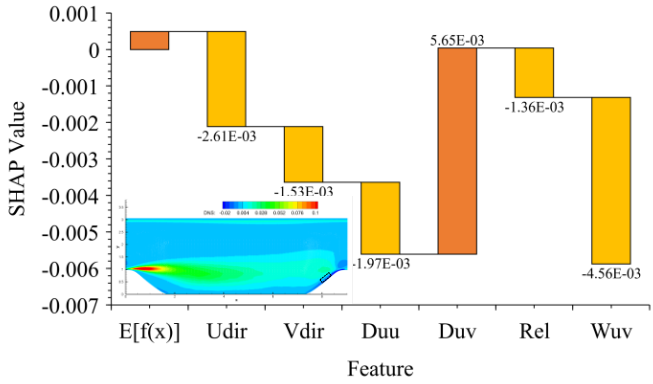
(f)

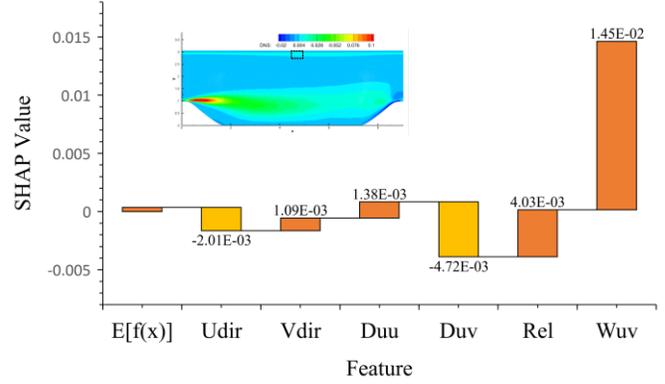
(g)

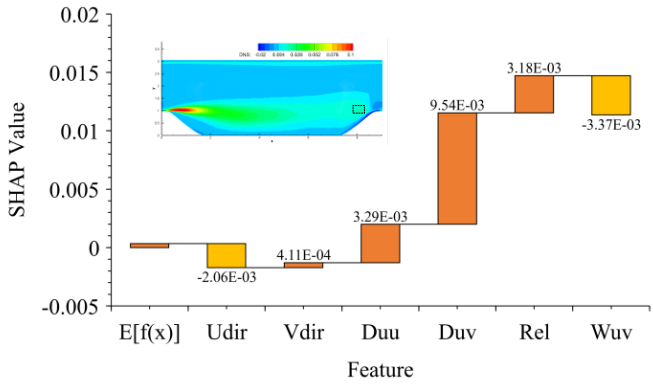
(h)

**Figure 15**: (a) SHAP value predictions at X = 0.75 (left column), 4.2 (middle column) and 6.5 (right column). The figure compares the machine learned $P_K$ predictions with DNS for the entire cross-section along with SHAP values of the different input features. Waterfall plot of the SHAP value predictions in the: (b) shear layer; (c) mid-channel; (d) backflow; (e) reattachment; (f) accelerating flow; (g) attached boundary layer; and (h) secondary peak regions. The region where the SHAP value is calculated is shown in each subfigure.

| %Hill Data | Correlation | Residual | Production (ML) |
|---|---|---|---|
| 14.8% | 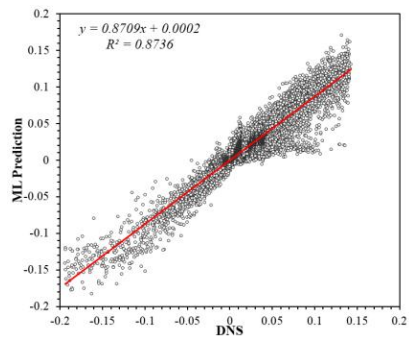 | 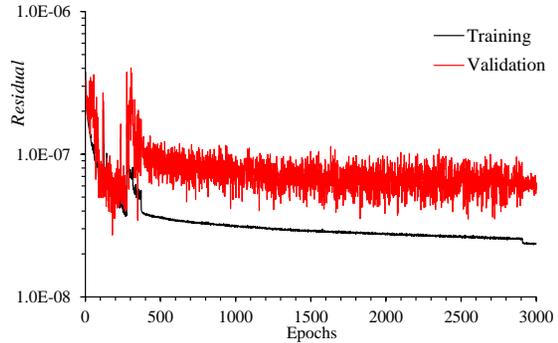 | 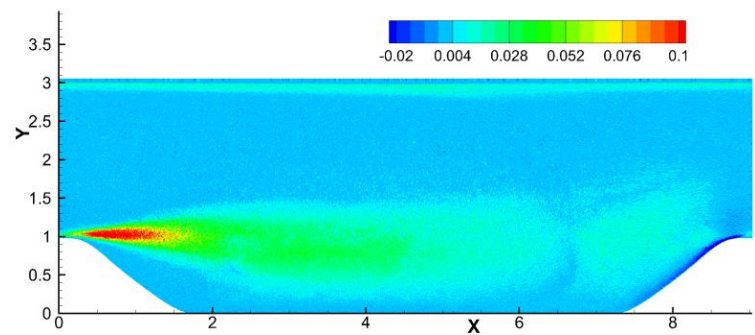 |
| 5.12% | 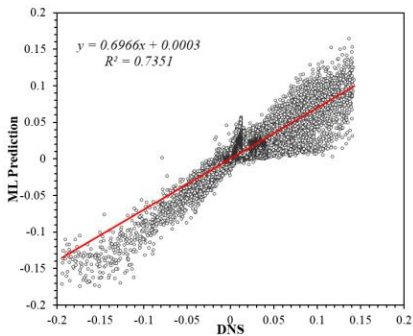 | 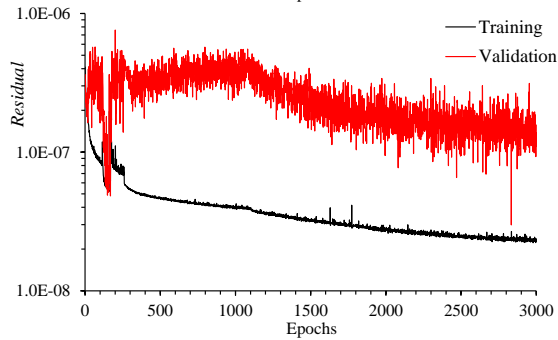 | 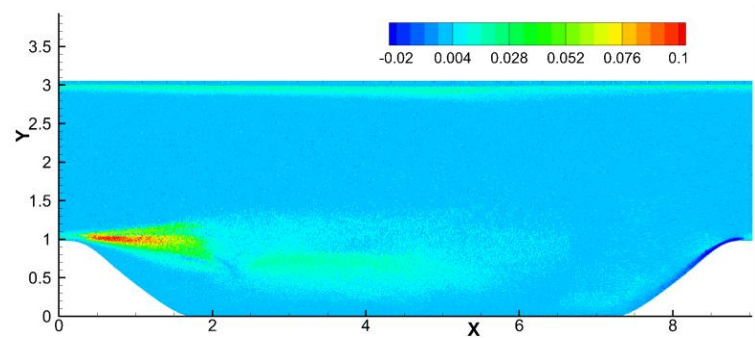 |
| 0% | 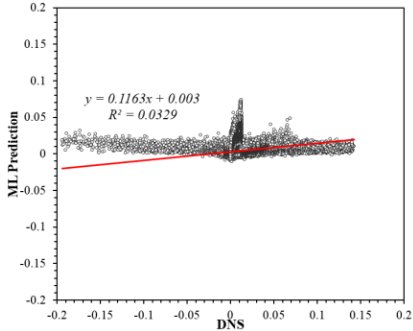 | 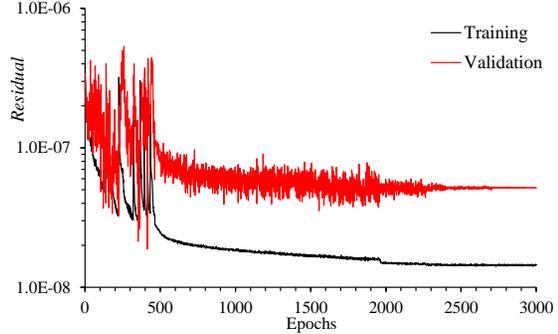 | 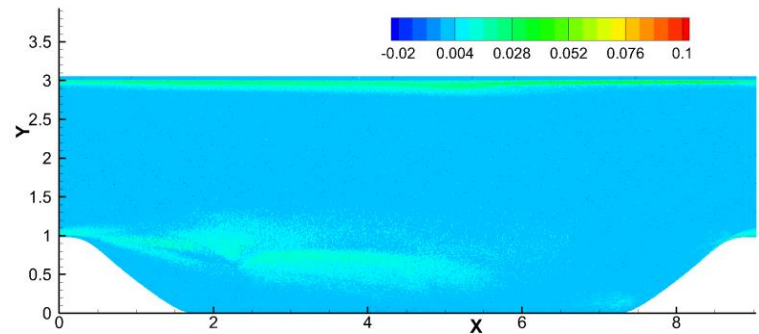 |

(a)

**Figure 16**: Continued.

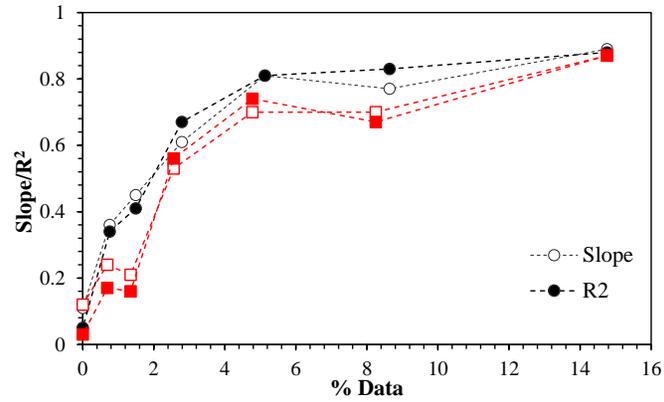

(b)

**Figure 16**: *A priori* tests of ML models. (a) Correlation between ML predictions and DNS datasets (left panel), variation of training and validation residual with epochs (middle panel), and contour plot of ML production prediction (right panel). Results are shown for ML model trained including $\frac{\bar{v}}{|u|}$. (b) Variation of $R^2$ and correlation line slope obtained when ML model predictions are compared with DNS. The abscissa shows the % of Hill data used during training. Black lines are for models trained including $\frac{\bar{v}}{|u|}$ as input feature, and Red lines are for models trained w/o $\frac{\bar{v}}{|u|}$.

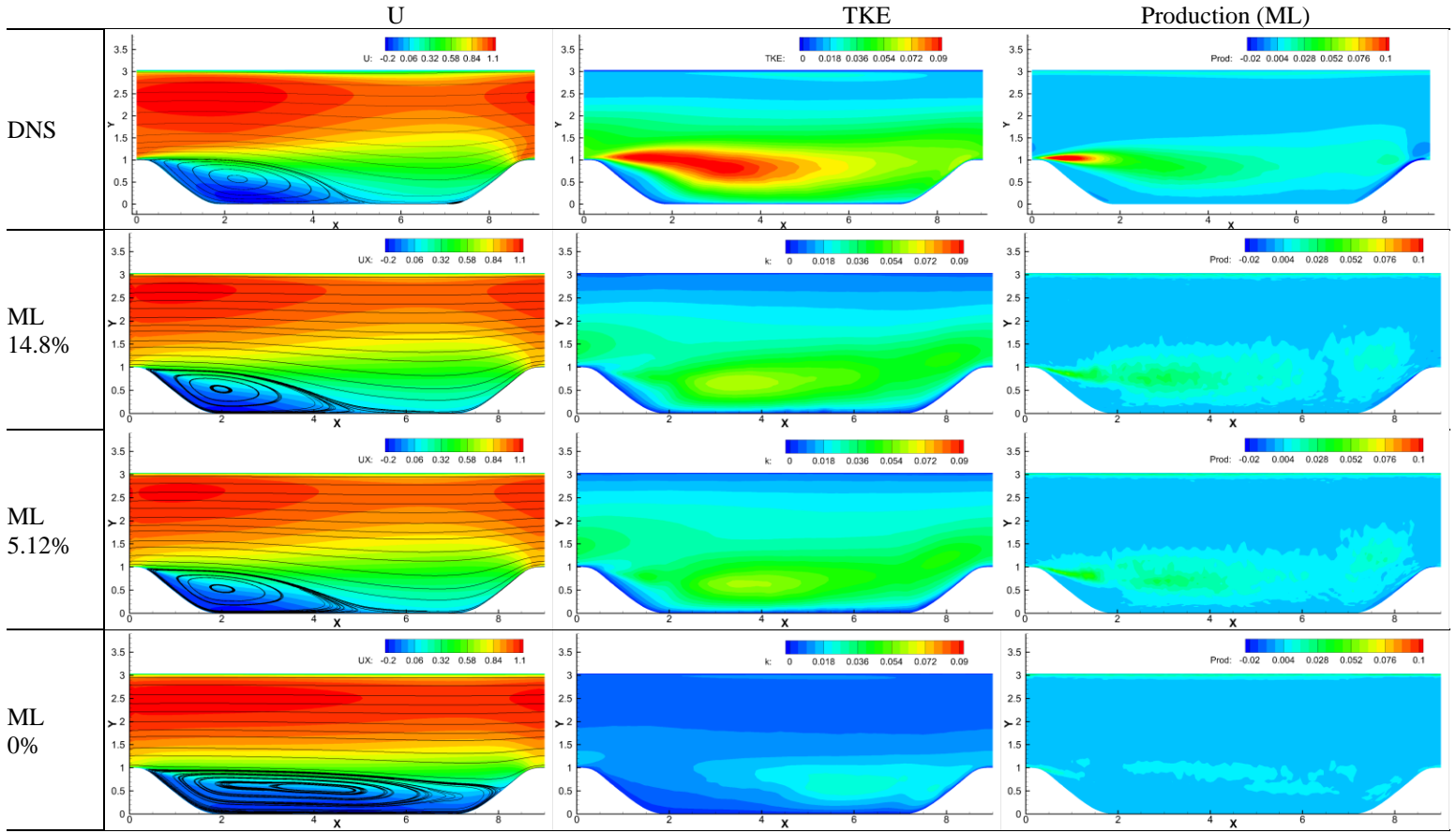

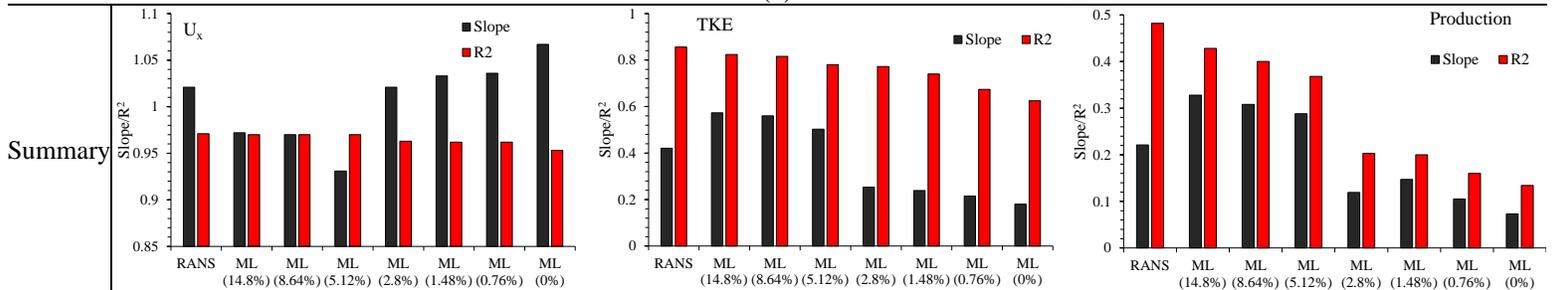

**Figure 17**: Results obtained using ML model in *a posteriori* tests are compared with DNS results. (a) The mean streamwise velocity and flow streamline (left column), TKE contour (middle column) and TKE production contour (right column) predicted by ML model trained using 15%, 5% and 0% of Hill data during training. (b) Summary plot showing the correlation between ML model predictions and DNS datasets for models trained using 15% to 0%. The plot also shows correlation between RANS results and DNS.

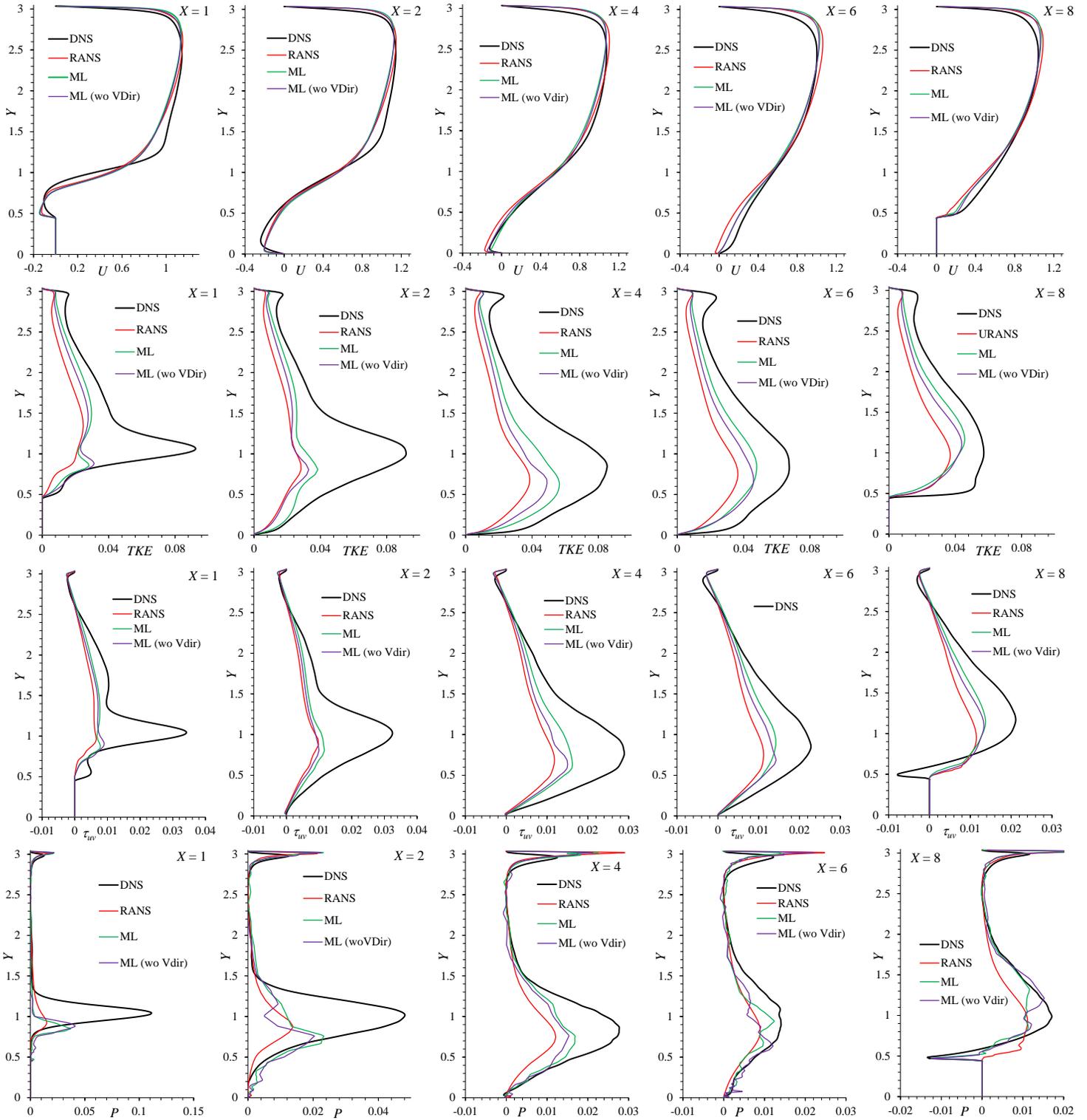

**Figure 18**: Comparison between ML model (with 5% Hill data) with DNS and RANS predictions on medium grid. The validation study compares mean streamwise velocity, TKE, shear stress and TKE productions at several cross-sections X = 1, 2, 4, 6 and 8.